\documentclass[traditabstract]{aa}
\usepackage{graphics}
\usepackage{amssymb, amsmath}
\usepackage{natbib}
\bibpunct{(}{)}{;}{a}{}{,}

\DeclareRobustCommand{\ion}[2]{%
\relax\ifmmode
\ifx\testbx\f@series
{\mathbf{#1\,\mathsc{#2}}}\else
{\mathrm{#1\,\mathsc{#2}}}\fi
\else\textup{#1\,{\mdseries\textsc{#2}}}%
\fi}

\def\apjl{ApJ}%
\def\apjs{ApJS}%
\def\ao{Appl.~Opt.}%
\def\aap{A\&A}%
\begin{document}

\title{Multi-scale Radio--IR Correlations in M~31 and M~33: \\
   The Role of Magnetic Fields and Star Formation}
\author{ F. S. Tabatabaei\inst{1}, E.\,M. Berkhuijsen\inst{2}, P. Frick\inst{3}
R. Beck\inst{2}, E. Schinnerer\inst{1} }
\institute{Max-Planck-Institut f\"ur Astronomie, K\"onigstuhl 17, 69117 Heidelberg, Germany
\and Max-Planck Institut f\"ur Radioastronomie, Auf dem H\"ugel 69, 53121 Bonn, Germany
\and Institute of Continuous Media Mechanics, Korolyov str. 1, 614013 Perm, Russia
  }
%
%
%

\titlerunning{ Radio--IR correlations in M~31 and M~33 }
\authorrunning{Tabatabaei et al.}

\abstract{Interstellar magnetic fields and the propagation of
cosmic ray electrons have an important impact on the
radio-infrared (IR) correlation in galaxies. This becomes evident
when studying  different spatial scales within galaxies. We
investigate the correlation between the infrared (IR) and free-free/synchrotron
radio continuum emission at 20\,cm from the two local group
galaxies M~31 and M~33 on spatial scales between 0.4 and 10\,kpc.
The multi-scale radio--IR correlations have been carried out using a
wavelet analysis. The free-free and IR emission are correlated on
all scales,  but on some scales the synchrotron emission is only
marginally correlated with the IR emission. The synchrotron--IR
correlation is stronger in M~33 than in M~31 on small scales
($<1$\,kpc), but it is weaker than in M~31 on larger scales.
{Taking the smallest scale on which the synchrotron--IR correlation
exists as the propagation length of cosmic ray electrons, we show
that the difference on small scales can be explained by the smaller propagation
length in M~33 than in M~31. On large scales, the difference is due to
the thick disk/halo in M~33, which is absent in M~31. A comparison of our
data with data on NGC 6946, the LMC and M~51 suggests that the propagation length
is determined by the ratio of ordered-to-turbulent magnetic field strength,
which is consistent with diffusion of CR electrons in the ISM. As the
diffusion length of CR electrons influences the radio--IR correlation, this
dependence is a direct observational evidence of the importance of
magnetic fields for the radio--IR correlation within galaxies. The star-formation
rate per surface area only indirectly influences the diffusion length as it
increases the strength of the turbulent magnetic field.} }

\keywords{galaxies: individual: M~33 -- galaxies: ISM -- galaxies:
magnetic field  -- ISM: cosmic rays -- infrared: galaxies -- radio
continuum: galaxies} \maketitle

%
%
\section{Introduction}

\label{sec:intro}
{  The correlation between radio and infrared continuum emission from galaxies is one of the tightest known in astronomy \citep[e.g.,][]{Helou_etal_85,deJong_etal_85,Gavazzi_etal_86}.  The global radio--IR correlation is nearly linear and invariant for over four orders of magnitude in luminosity \citep[e.g. ][]{Yun} and holds for all kinds of star forming galaxies \citep[see][and references therein]{Condon_92}. The linearity of the correlation is conventionally attributed to a common dependence of IR and radio emission on massive star formation. Massive stars heat the dust and ionize the gas leading to thermal dust emission and thermal (free-free) radio emission. The same stars are progenitors of supernova remnants (SNRs) that produce cosmic ray electrons (CREs) radiating synchrotron emission. However, an important omission in this simple picture is the         role of magnetic fields that determine the intensity of the synchrotron emission with about the square of the field strength. \cite{Lacki_10} modelled all relevant processes and found that the linearity of the global radio-IR correlation is  due to a conspiracy of several factors \citep[see also][]{Bell}. For instance, in galaxies of low surface  brightness and low star formation rate, the weak IR emission is balanced by weak radio emission caused by escape of CREs.

The radio-IR correlation also holds within galaxies, where local variations were found by several authors 
\citep[e.g.][]{Beck_88,Hoernes_etal_98,Gordon_04,Hughes_etal_06,Dumas,Taba_13}}.
Detailed multi-scale analysis showed that the smallest scale on
which the radio--IR correlation occurs is not the same in the LMC
\citep[][]{Hughes_etal_06} and M~51 \citep[][]{Dumas}, the reason
being unclear. Variations may be caused by differences in massive SF
or dust heating sources,  in magnetic field structure or in the
propagation of cosmic ray electrons (CREs).

One fact making the IR--SF linkage less obvious is that the IR
emission  consists of at least two emission components, cold dust
and warm dust. The cold dust emission may not be directly linked to
the young stellar population as  it is heated by the interstellar
radiation field \citep[ISRF,][]{Xu_90}. Furthermore, it is necessary
to distinguish between the two main radio continuum (RC) components,
free-free and synchrotron emission.  The free-free emission from
electrons in HII regions around young, massive stars is expected to
be closely connected to the warm dust emission that is heated by the
same stars \citep[e.g.][]{Condon_92_1}. Although the CREs responsible for synchrotron emission
also originate from SF regions (supernova remnants, the final
stages in the evolution of massive stars), the synchrotron--IR
correlation may not be as tight as the free-free--IR correlation
locally, as a result of the propagation of CREs from their
places of birth \citep[see][and references therein]{Longair}. The CREs experience various energy losses
interacting with matter and magnetic fields in the ISM, causing
variation in the power law index of their energy distribution. A
significant variation was found in M~33 by using a thermal radio tracer (TRT) method
separating free-free and synchrotron emission
that revealed flatter synchrotron spectra in
regions of massive SF than in between the arms and in the outer
disk \citep{Tabatabaei_3_07}. This method yields more realistic free-free and synchrotron
maps than the conventional separation based on a constant
synchrotron spectral index.

As the variations in the radio--IR correlation are possibly due to a
range of different conditions such as  the SF rate (SFR), heating
sources of dust, magnetic fields, and CRE propagation, {  one
needs to separate the free-free and synchrotron contributions to the RC emission in
nearby galaxies that allow detailed, well resolved studies. Such a detailed study has been
recently carried out in NGC\,6946, which showed, for the first time, a direct link between the
variation of the radio-IR correlation and the total magnetic field strength \citep{Taba_13}.
In the present} paper, we separate the free-free and synchrotron emission from the
Andromeda galaxy using the same  TRT method as for M~33
{  \citep{Tabatabaei_3_07}}. The influence of different star formation
activities and ISM conditions on the radio--IR correlation is
ideally traced by comparing early-type and late-type spirals like
the nearest galaxies M~31 \citep[D=\,780\,kpc,][]{Stanek} and M~33
\citep[D=\,840\,kpc,][]{Freedman_etal_91}. {  The free-free and synchrotron emission maps of }
M~33 and M~31 enable us to {  separately} compare the free-free--IR
correlation and synchrotron--IR correlation in these galaxies.

In this paper, we analyze the radio--IR correlation
scale-by-scale rather than using classical pixel-by-pixel
correlations. The usual pixel-by-pixel correlation between the
distributions of FIR and radio emission contains all scales that
exist in a galaxy image. For example, the high-intensity points
represent high-emission peaks {at} small scales belonging to bright
sources, whereas low-intensity points generally represent weak
diffuse emission with a large-scale distribution.


A scale-by-scale approach is required for an unbiased recognition
of similarities and differences {  between images tracing different emission mechanisms.}
Such a study enables us to uncover the scale dependence of
the correlation of emitting structures. This method helps to identify the smallest scale where the radio--FIR correlations break
down, and provides important information on the propagation length of
CREs.

To separate the contribution { from} different spatial scales in the global
correlation of two maps \cite{Frick_etal_01} introduced the `wavelet
cross--correlation'. The wavelet transform, similar to the Fourier
transform, decomposes the images scale-by-scale, but, in contrast to
the Fourier decomposition, wavelets keep the information about the
distribution of structures of a given scale in the physical space
and therefore allow one to analyse the correlation of two fields
scale-by-scale.

We describe the relevant data sets in Sect.\,2. The { separation
of the RC maps in free-free and synchrotron components in M~31 is
explained} in Sect.\,3. Wavelet decompositions and wavelet spectra of
the radio and IR emission, and their cross correlations are
presented in Sect.\,4. We discuss and summarize our results in
Sect.\,5 and 6, respectively.
{
\begin{table*}
\begin{center}
\caption{Images of M~33 and M~31 used in this study. }
\begin{tabular}{ l l l l l}
\hline
 &\,\,\,\,\,\,\,\,\,\,\,\,\,\,\,\,\,\,\,\, M\,31& &\,\,\,\,\,\,\,\,\,\,\,\,\,\,\,\,\,\,\,\, M\,33 &\\
Wavelength & Resolution &  Telescope  & Resolution &  Telescope  \\
\hline
20\,cm       &  $45\arcsec$ & VLA+Effelsberg $^{1}$ & $51\arcsec$ &  VLA+Effelsberg $^{2}$\\
160\,$\mu$m  &  $40\arcsec$ &Spitzer-MIPS$^{3}$ &$40\arcsec$ &Spitzer-MIPS$^{4}$ \\
70\,$\mu$m  &  $18\arcsec$  &Spitzer-MIPS$^{3}$ & $18\arcsec$ &Spitzer-MIPS$^{4}$\\
24\,$\mu$m  &  $6\arcsec$  &Spitzer-MIPS$^{3}$ & $6\arcsec$ &Spitzer-MIPS$^{4}$\\
6563\AA{}\,(H$\alpha$)   &$1.5\arcsec$ &  KPNO$^{5}$ &$2\arcsec$ & KPNO$^{6}$\\
\hline
\noalign {\medskip}
\multicolumn{3}{l}{$^{1}$ \cite{Beck_98}}\\
\multicolumn{3}{l}{$^{2}$ \cite{Tabatabaei_3_07}}\\
\multicolumn{3}{l}{$^{3}$ \cite{Gordon_06}}\\
\multicolumn{3}{l}{$^{4}$ \cite{Tabatabaei_1_07}}\\
\multicolumn{3}{l}{$^{5}$ \cite{Devereux_etal_94b}}\\
\multicolumn{3}{l}{$^{6}$ \cite{Hoopes_et_al_97H}}\\
\end{tabular}
\end{center}
\end{table*}
}
\section{ Data}
\label{sec:data}
Table 1 summarizes the data used in this work. The interferometeric observations of M~31 and M~33 at 20\,cm were
made with the Very Large Array (VLA\footnote{The VLA (Jansky VLA) is
a facility of the National Radio Astronomy Observatory. The NRAO is
operated by Associated Universities, Inc., under contract with the
National Science Foundation.}). Single-dish observations at the same
wavelength were carried out with the 100-m Effelsberg telescope\footnote{The 100--m telescope at Effelsberg is operated by the
Max-Planck-Institut f\"ur Radioastronomie (MPIfR) on behalf on the
Max--Planck--Gesellschaft.}. We used the combined VLA + Effelsberg
20\,cm radio continuum data of \citet{Beck_98} for M~31 and of
{ \citet{Tabatabaei_2_07}} for M~33.

We used the Spitzer MIPS data at 24\,$\mu$m, 70\,$\mu$m, and
160\,$\mu$m for M~31 \citep{Gordon_06} and M~33
\citep{Tabatabaei_1_07}.  The basic data reduction and mosaicing was
performed with the MIPS instrument Data Analysis Tool Version 2.90
\citep{Gordon_05}. For both galaxies, the sky and background sources in the MIPS maps were subtracted as detailed in \cite{Gordon_06} and \cite{Tabatabaei_1_07}.

The H$\alpha$ observations of M~31 were carried out on the Case
Western  Burrell--Schmidt telescope at the Kitt Peak National
Observatory, covering a field of view of
$2^{\circ}\,\times\,2^{\circ}$  for M~31 \citep{Devereux_etal_94b}
and of $68\arcmin$\,$\times$\,$68\arcmin$ for M~33
\citep{Hoopes_et_al_97H}. Here, we use the de-reddened H$\alpha$
maps of M~31 and M~33 presented in \cite{Tabatabaei_10} and
\cite{Tabatabaei_3_07}, respectively.

For M~31, we subtracted background radio sources with flux densities
$S\,>$35\,mJy from the combined 20\,cm radio map{\footnote{  Any point source 
brighter than 35\,mJy  in M~31 is too bright to be an SNR.}.  Furthermore, the
bright nucleus of M~31 was subtracted from all radio and IR maps
before the wavelet analysis, as it causes a strong bias in our
wavelet results. For M~33, all known background radio sources given
by \cite{Viallefond} were subtracted. The MIPS maps were convolved
to the resolution of the VLA 20\,cm radio data (45\arcsec\ for M~31
and 51\arcsec\ for M~33) by using the custom kernels presented by
\citet{Gordon_07}. After convolution, the maps were normalized in
grid size, orientation and reference coordinates.

\section{Free-free and synchrotron emission}
\label{free}
We separate free-free and synchrotron emission from M\,31 and M\,33
using the TRT method in which one of the hydrogen recombination lines is used as a template for the free-free emission \citep[see e.g. ][]{Dickinson,Tabatabaei_3_07}. The brightest recombination line data available,  the H$\alpha$ line emission, is used for both galaxies. This pixel-by-pixel separation  results in maps of the free-free and synchrotron emission for M~31 and M~33.
\subsection{M~31}
\label{sec:temission}
We use the de-reddened H$\alpha$ emission presented in
\cite{Tabatabaei_10} and convert it to the emission measure via the
expression of \cite{Valls}
\begin{equation}
I_{{\rm H}\alpha} = 9.41 \cdot 10^{-8} \, T^{-1.017}_{e4} \,
10^{-(0.029/T_{e4})} \, {\rm EM} \ ,
\end{equation}
where the electron temperature, $T_{e4}$, is in units of $10^4$\,K,
EM in cm$^{-6}$ pc, and it is assumed that the optical depths of HI
resonance lines are large (usually denoted as case B). Optical
spectroscopic observations of the HII regions in M~31 indicate that
the electron temperature ($T_e$) increases with galactocentric
radius as a result of the metallicity gradient \citep{Dennefeld,
Blair}. Figure~1 shows the $T_e$ measurements
of these authors  obtained from the ([OII] +[OIII])/H$\beta$
ratios. A least-squares fit to their results gives
%
\begin{equation}
\label{eq:Te}
T_{e4} = (0.017\,\pm\,0.005)\,R + (0.5\,\pm\,0.1),
\end{equation}
with $R$ the galactocentric radius in kpc,  scaled to
D=780\,kpc and corrected for the inclination of 75$^{\circ}$ before
fitting the above line. The Galactic measurements of \cite{Madsen}
give an average electron temperature $T_{e}$ of about 1000\,K higher
in diffuse ionized gas (DIG) than in HII regions. Assuming that the
same  difference occurs in M~31, and taking an HII region (DIG)
contribution of 60\%(40\%) to the total H$\alpha$ emission
\citep[e.g.][]{Greenawalt}, we derived the EM from Eqs.~(1) and (2).
The EM is further related to the continuum optical thickness,
$\tau_c$, of the ionized gas by
\begin{equation}
\tau_c = 8.235 \cdot 10^{-2} \, a \, T_e^{-1.35} \, \nu_{{\rm
GHz}}^{-2.1} \, (1+0.08) \, {\rm EM} \ ,
\end{equation}
with $a \simeq 1$ \citep{Dickinson}. The factor (1\,+\,0.08) takes
into account the contribution from  singly ionized He. The
brightness temperature of the radio continuum emission, $T_b$, then
follows from
\begin{equation}
T_b = T_e \, (1-e^{-\tau_c}) \ .
\end{equation}
\begin{figure*}
\begin{center}
\resizebox{7cm}{!} {\includegraphics*{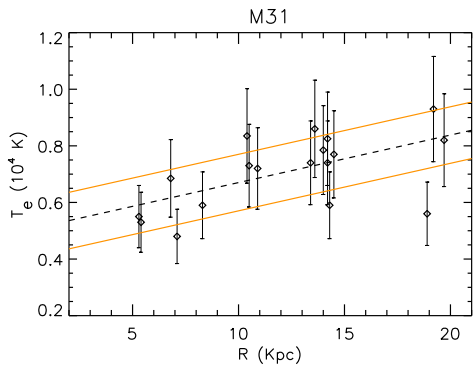}}
\caption[]{The electron temperature estimates for M~31, based on the
empirical method, the ([OII] +[OIII])/H$\beta$ ratio, by
\cite{Dennefeld} and \cite{Blair} after correcting for M~31's
inclination of 75$^{\circ}$ and scaling to D=780\,kpc. Also
shown are the fitted least-squares slope (dashed line) and the
5\,$\sigma$ confidence levels (red solid lines). }
\end{center}
\label{fig:temp}
\end{figure*}
%
%
The resulting distribution of the intensity of the predicted
free-free emission in Jy/beam at 20\,cm is shown in Fig.~2 (top panel). Strong
thermal emission emerges from the center of the galaxy with a
maximum intensity of $\simeq$\,4\,mJy/beam and from complexes of
star-forming regions that are most abundant in the north-east of the
star-forming ring (the so-called `10\,kpc ring',
$30\arcmin<R<55\arcmin$).  By integrating the maps of the observed radio continuum
and the free-free emission in rings around the galaxy center out to
a radius of 16\,kpc, we obtain the thermal flux density and thermal
fraction (Table~2). The thermal fraction is 15\%\,$\pm$\,2\%
in the  `10\,kpc ring' and 30\%\,$\pm$\,4\% for $R<30\arcmin$
(6.8\,kpc).
The thermal flux density of $S_{\rm th}=630\pm60$\,mJy is lower than the value of $0.8\pm0.1$\,Jy obtained by \cite{Hoernes_etal_98}, who used a constant nonthermal spectral index to separate the thermal and nonthermal components. A similar overestimate of the thermal emission was found in M~33 assuming a constant nonthermal spectral index \citep[the so called standard thermal/nonthermal separation method,][]{Tabatabaei_3_07}. This is {  due to the synchrotron emission being underestimated} when the nonthermal spectral index is flatter than the one assumed in the standard method. 

Possible origins of the strong H$\alpha$ emission and high thermal
fraction at $R<30\arcmin$ have been discussed by
\cite{Devereux_etal_94b} and \cite{Tabatabaei_10}. As recent SF is
scarce in this region, the gas must have been ionized by shocks, evolved
stars  \citep[][]{Groves_12} and/or during the last encounter with a companion galaxy
\citep{Block}.

\begin{table}
\begin{center}
\label{table:scalelength} \caption{Thermal fractions of the radio continuum emission, $F_{\rm th}$, in M~31 at 20\,cm .}
\begin{tabular}{ l l l l l}
\hline
$R$ &  $R$ & $S_{\rm obs}$ &  $S_{\rm th}$ & $F_{\rm th}$ \\
(kpc) & ($\arcmin$)    &  (mJy) & (mJy) & \% \\
\hline
0-16 & 0-70    & 4000\,$\pm$\,400   &  630\,$\pm$\,60   &  16\,$\pm$\,2  \\
6.8-12.5 & 30-55  & 2600\,$\pm$\,300    & 380\,$\pm$\,40   &  15\,$\pm$2  \\
0-6.8 &  0-30     & 660\,$\pm$\,70      &  200\,$\pm$\,20  & 30\,$\pm$\,4 \\
0-1   & 0-4.4   & 120\,$\pm$\,10      &  31\,$\pm$\,3      & 26\,$\pm$\,3 \\
\hline
\end{tabular}
\tablefoot{ Integrated flux density of the
total ($S_{\rm obs}$) and thermal ($S_{\rm th}$) radio continuum emission at 20\,cm for various radial intervals in the galactic plane (inclined by 75$^{\circ}$). A
distance of 780\,kpc was used for these calculations. Errors are
systematic errors due to uncertainty in the zerolevel.}
\end{center}
\label{tab:flux}
\end{table}

\begin{figure*}
\begin{center}
\resizebox{14cm}{!}{\includegraphics*{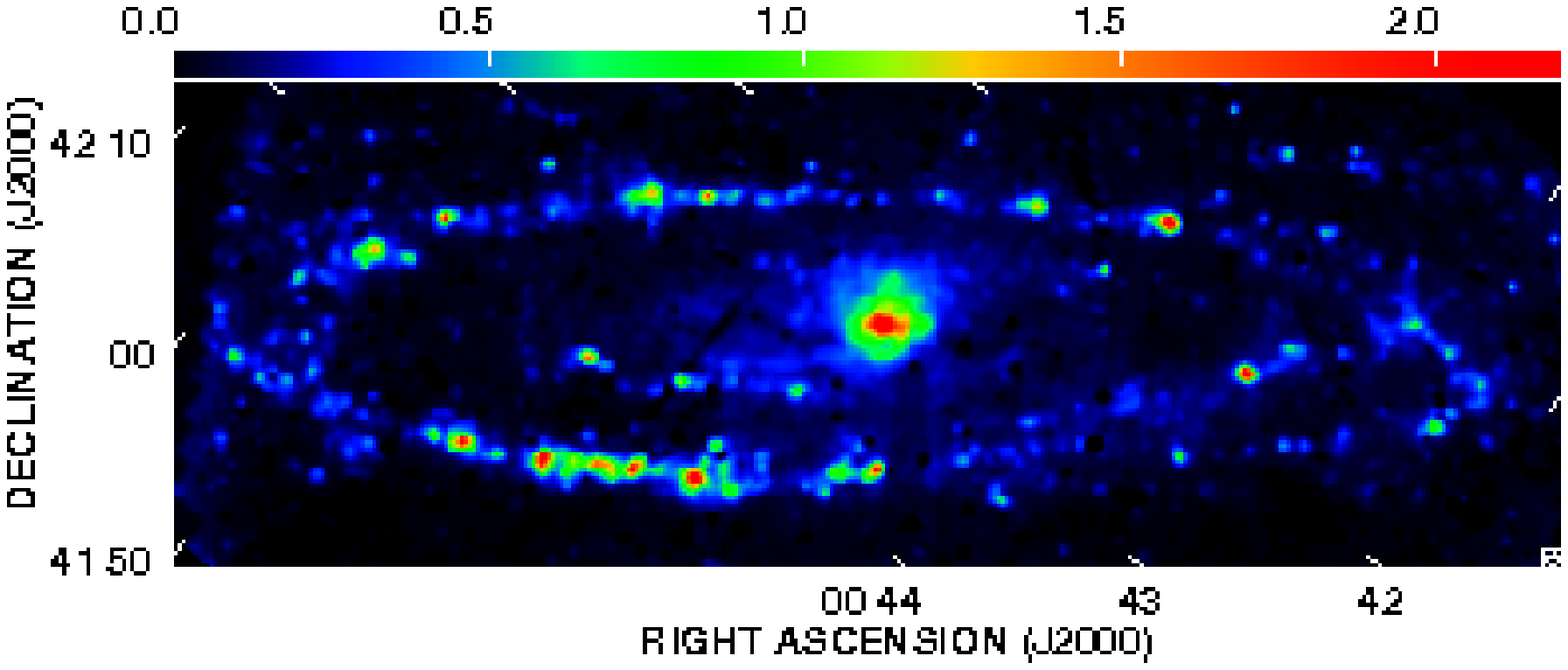}}
\resizebox{14cm}{!}{\includegraphics*{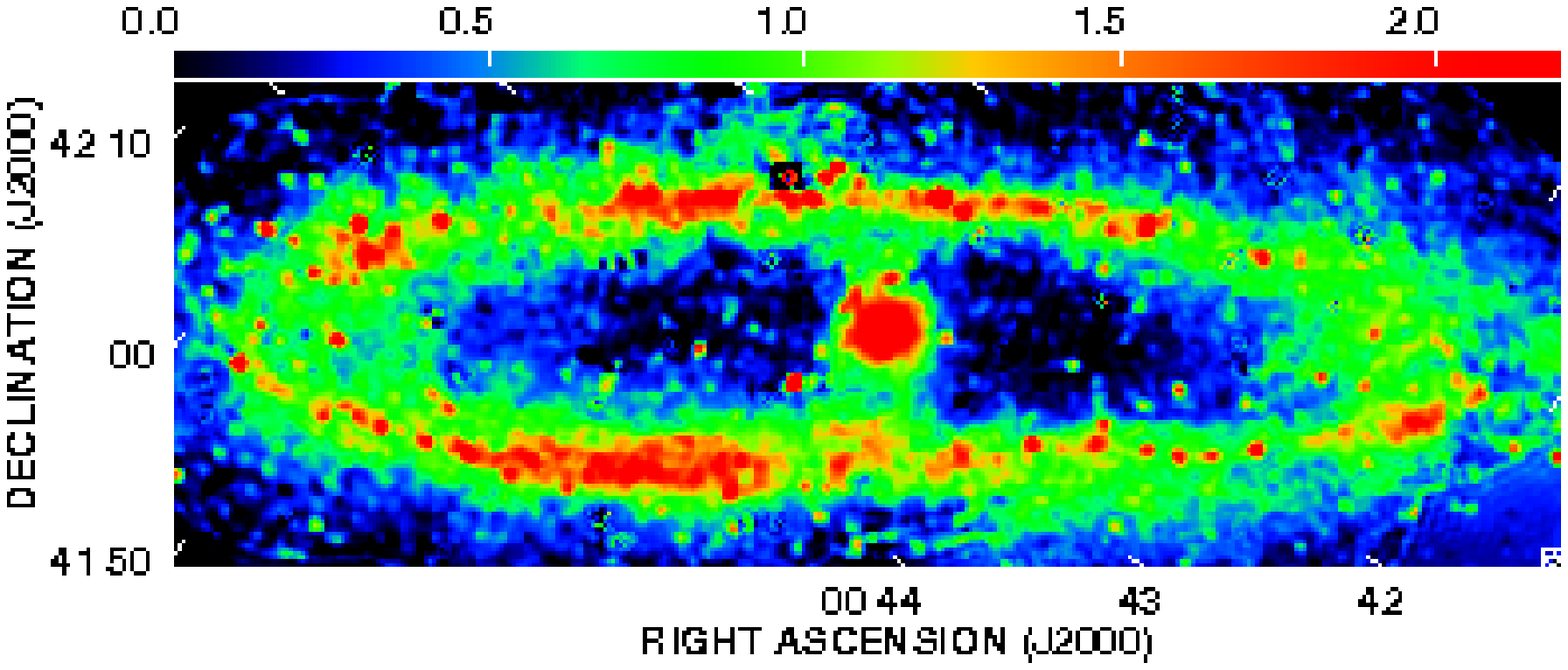}}
\caption[]{ Thermal free-free ({\it top}) and nonthermal synchrotron  emission ({\it bottom}) from M~31 at 20\,cm, separated using the TRT method \citep{Tabatabaei_3_07}. The color scale gives the flux density in mJy/beam. The angular
resolution is 45$\arcsec$ (shown in the lower right of the images)  corresponding to $\simeq$ 180\,pc linear resolution along the major axis. }
\end{center}
\label{fig:m31th20}
\end{figure*}
Subtracting the free-free emission from the observed radio emission
at 20\,cm, we derive the distribution of the synchrotron emission
(Fig.~2, bottom). This map exhibits a wide ring of diffuse
emission coinciding with  the `10\,kpc ring'. Interestingly, the
synchrotron emission is stronger close to  star forming regions.
Some of these regions are accompanied by patches of relatively
high-intensity ($\gtrsim$\,2\,mJy/beam) synchrotron radiation.
The strongest synchrotron
emission emerges from the center of the galaxy with a maximum of
9\,mJy/beam. The extended, diffuse structure around the spiral
arms indicates that CREs propagated away from their places of
origin.

The high thermal fraction of about 30\% at $R<30'$ (6.8\,kpc)  is partly due to the weak synchrotron emission from regions
outside the central region ($2.5<R<4.5$\,kpc) with a minimum of
240\,$\mu$Jy/beam (about 3 times the noise level). This is the
result of a
lack of cosmic ray electrons, since there are magnetic fields but
there is very little
recent star formation in this region \citep{Moss,Berkhuijsen_89}.
\subsection{M~33}
%
%
\begin{figure*}
\begin{center}
\resizebox{\hsize}{!}
{\includegraphics*{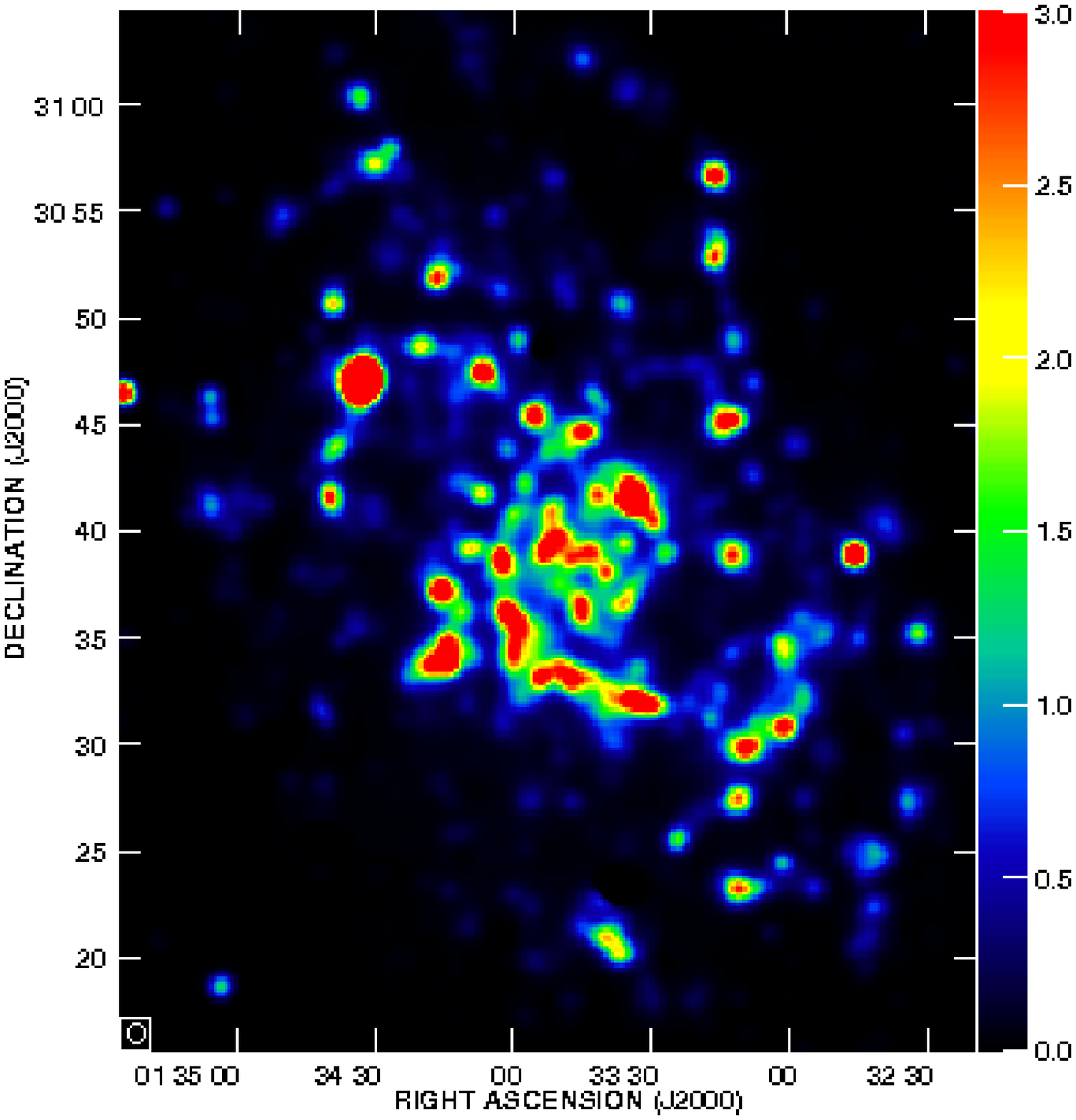}
\includegraphics*{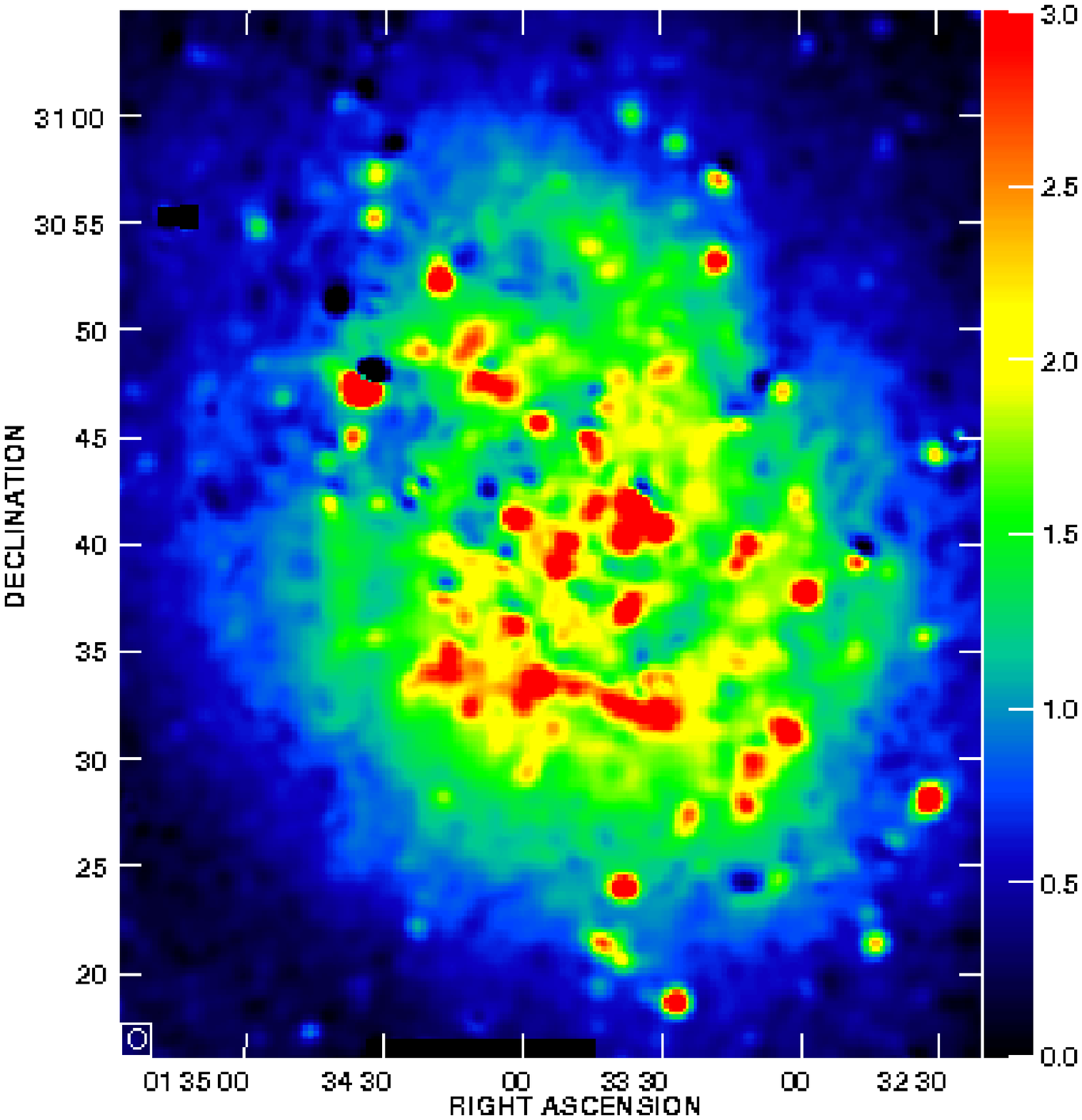}}
\caption[]{ Thermal ({\it left}) and synchrotron ({\it right}) radio
continuum emission from M~33 at 20\,cm at 51$\arcsec$ resolution,  separated using the TRT method \citep{Tabatabaei_3_07}. The color scale gives the flux density in mJy/beam.}
\end{center}
\label{fig:m33nt20}
\end{figure*}
We presented the free-free and synchrotron maps of M~33 in
\cite{Tabatabaei_3_07} at $90\arcsec$ angular resolution. In this
paper, for a proper comparison with M~31, we re-derive the free-free
and synchrotron maps at the resolution of the 20\,cm VLA map,
$51\arcsec$ {  \citep{Tabatabaei_1_07}}  (Fig.~3).  The free-free map is
dominated by HII regions, which are better resolved at $51\arcsec$
resolution compared to the results reported by \cite{Tabatabaei_3_07}.
Moreover, smooth synchrotron spiral arms are detected with
intensities of  $\lesssim$\,2\,mJy/beam ($\simeq 10\sigma$), which were
not visible  at $90\arcsec$ resolution due to beam smearing.
The strong synchrotron clumps close to the sites of active star
forming regions shown in \cite{Tabatabaei_3_07} exhibit several
sub-structures at $51\arcsec$ resolution. The general distributions of the synchrotron
emission and the thermal fraction, however, remain the same as  before.
Most of the distinct point sources coincide with supernova remnants,
as shown in \cite{Tabatabaei_3_07}.
\begin{figure*}
\begin{center}
\resizebox{9cm}{!}{\includegraphics*{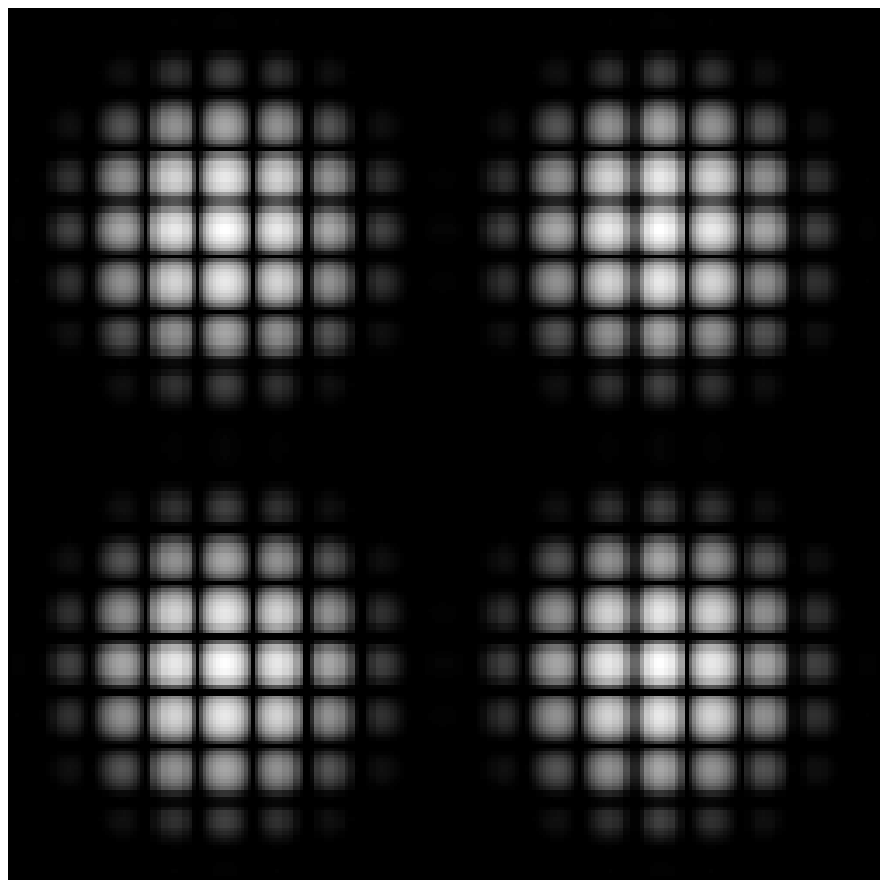}
\includegraphics*{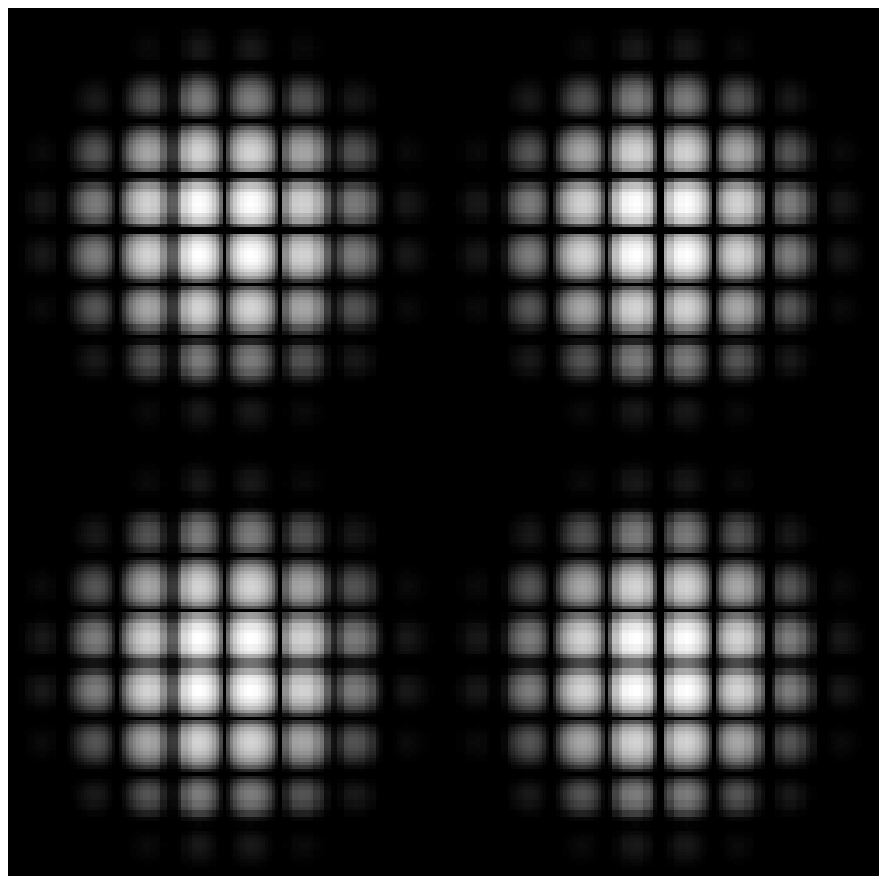}}
\resizebox{12cm}{!}{\includegraphics*{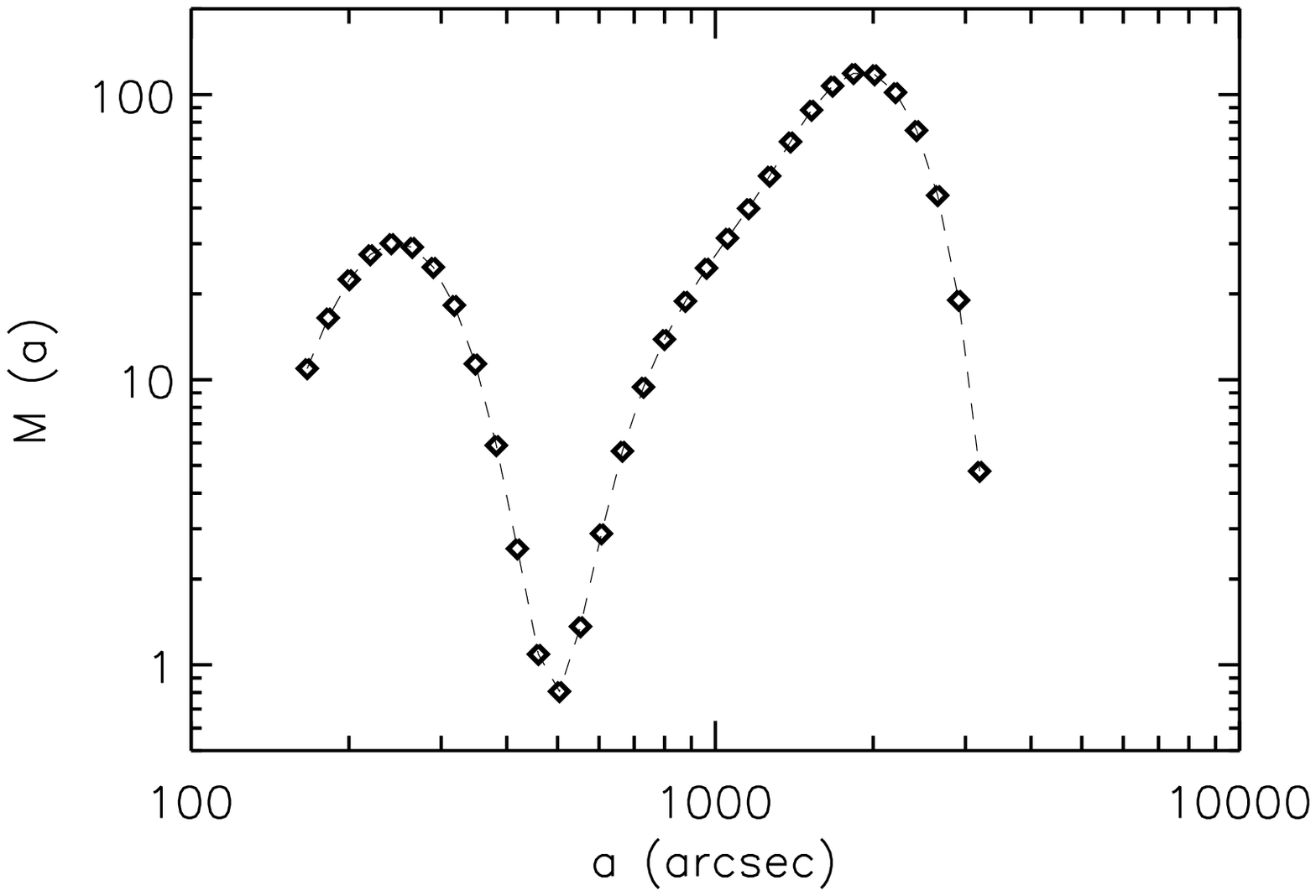}
\includegraphics*{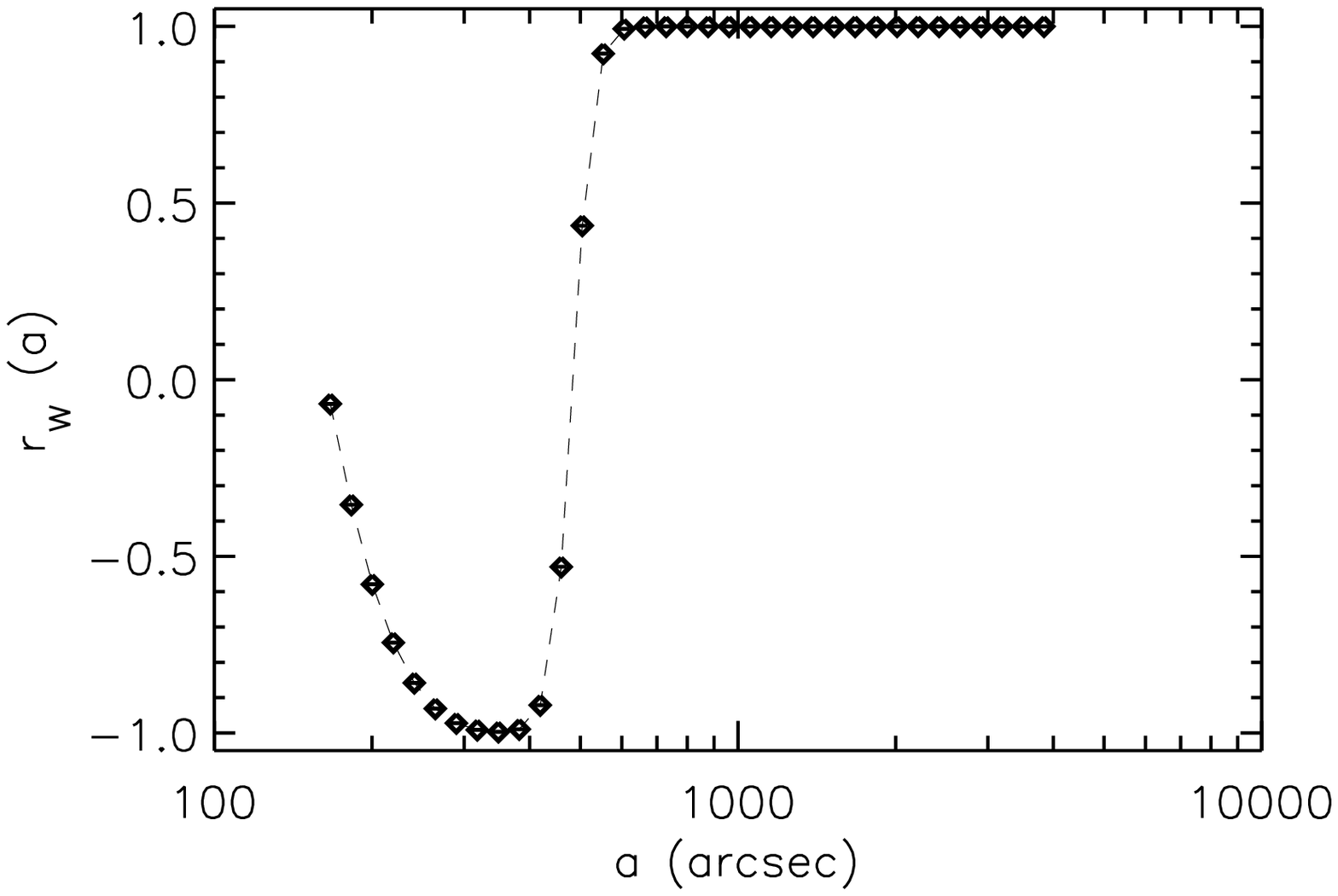}}
\caption[]{An artificial example of a scale-by-scale wavelet
cross-correlation analysis. {\it Top}: {  two different mock maps before the wavelet decomposition. The maps show four large-scale spots made by small-scale spots.} The large-scale spots in the two maps  slightly differ in size (6 rows and columns in the right panel, 5 rows and columns in the left panel). As they are centered at the same location, the small-scale spots are shifted with respect to each other.
{\it Bottom left}: wavelet spectra (they are identical for both maps),
{\it bottom right}: wavelet cross-correlation function. }
\label{fig:test}
\end{center}
\end{figure*}
\section{Wavelet analysis and scale-by-scale correlation}
The next step of our analysis is the scale-by-scale
comparison of the radio and IR emission from M~31 and M~33. We first describe the
basics of the wavelet decomposition, then apply this method to decompose the radio and IR
emission from each galaxy.
\subsection{Algorithms}
\label{algorithms}
 To illustrate why a
scale-by-scale analysis can be {  useful} we present in
Fig.~\ref{fig:test} an example. Two artificial maps
are shown in this figure, both include four large-scale spots
formed by a set of small-scale spots. The large-scale spots are
located at the same coordinates, but the small-scale spots in one
map are shifted to in between the small-scale spots in the other
map. The two maps look very similar, but their cross-correlation
$$
c=\frac{\int (f_1({  x}) - \overline{f_1})(f_2({  x}) - \overline{f_2})d{  x}}
{\left(\int (f_1({  x}) - \overline{f_1})^2d{  x} \int (f_2({  x}) - \overline{f_2})^2d{  x}
\right)^{1/2}}=0.03$$
is weak (here $f_1$ and $f_2$ are the intensities of both maps, {  and $\overline{f_1}$ {  and  $\overline{f_2}$} denote averages. }

For this simple example it is obvious that the large scales are
completely correlated and the small scales are anti-correlated.
For less obvious cases one needs some formal algorithm to separate
different levels of correlation on different scales. The algorithm
should include two steps: first, some scale filtering, which
isolates the structures in a given scale range, and second,
calculation of correlation in filtered maps. We note that
{  Fourier filtering is of limited use in this situation} because each Fourier
mode covers the whole physical space and leaves no possibility for
statistics over the map. Instead, one needs a scale filtering
which keeps the independency of structures in physical space, i.e.
filtering, localized in both Fourier and physical space. {  This is the underlying} idea of the {\it wavelet transform} which
makes a spatial--scale decomposition using the convolution of the
data with a family of self-similar basic functions that depend on
the scale and the location of the structure. The wavelet
coefficients for the 2D continuous wavelet transform are given by:
\begin{equation}
W(a,{  x})=\frac{1}{a^{\kappa}} \int_{-\infty}^{+\infty} \int_{-\infty}^{+\infty} f(  x')\psi^{\ast}(\frac{{  x'-x}}
{{\it a}}){\it d}{  x'},
\label{wav_transform}
\end{equation}
\noindent where $f({  x})$ is the analyzed, two--dimensional
function (the image), $\psi({  x})$ is the analyzing wavelet, the
symbol $^{\ast}$ denotes the complex conjugate, ${  x} = (x,y)$
defines the position of the wavelet, {\it a} defines its scale, and $\kappa$ is the normalization parameter.
In Fourier space, the wavelet coefficients (Eq.~5) can be expressed as
\begin{equation}
W(a,{  x})=\frac{a^{2-\kappa}}{4\pi^2} \int_{-\infty}^{+\infty} \int_{-\infty}^{+\infty} \hat{f}(  k)\psi^{\ast}({\it a  k}) {  e}^{ikx} {\it d   k},
\label{wav_transform2}
\end{equation}
\noindent where $\hat{f}(  k)$ is Fourier transform of the function $f(  x)$ and $  k=(k_x,k_y)$ is the wavevector. Eqs. (\ref{wav_transform}) and (\ref{wav_transform2}) allow one to plot a 2D map for a given scale $a$ (to be more precise, for a range of scales $a\pm\Delta
a/2$, where $\Delta a$ is the scale resolution of the analyzing
wavelet).

The choice of the analyzing wavelet $\psi$  is very
important and strongly depends on the goal of the performed
analysis. For the scale-by-scale cross-correlation analysis one
needs a good scale resolution $\Delta a$ (to separate the
contribution from different spatial scales) and a good spatial resolution
$\Delta x$ (to increase the number of independent areas). These two
requirements are {  mutually exclusive} ($\Delta x \, \Delta a = 2 \pi$). A
reasonable compromise for the cross-correlation analysis of 2D maps
has been suggested by \citet[][]{Frick_etal_01}, who used the
wavelet function called `Pet Hat', first introduced in
\cite{Aurell} for turbulent flow modeling. The Pet--Hat function
is defined in Fourier space by the formula
\begin{equation}
\psi(  k)=\left \{ \begin{array}{ll}
{\cos^2\Big(\frac{\pi}{2}\,\log_{2}\,(\frac{k}{2\pi})\Big)}  & \,\,\,\,\,\pi \le k  \le 4\pi \\
0 &   k  < \pi  \, \,\, {\rm or} \,\,\,  k  > 4\pi , %
\end{array} \right.
\label{pethat}
\end{equation}
\noindent where ${  k}$ is the wavevector and $k=\vert   k \vert$. This
function has been used by several authors to study the scale
distribution of various {  emission mechanisms} in galaxies
\citep[e.g.][]{Frick_etal_01,Hughes_etal_06,Laine,Dumas}.  2D
continuous Pet--Hat transformations of the Spitzer MIPS maps at 24,
70, and 160\,$\mu$m were presented for M\,33 by \cite{Tabatabaei_1_07} and for M\,31 by \cite{Tabatabaei_10}.

Similar to the standard Fourier energy spectrum given by $E(k)=\int_{\vert   k\vert} \vert \hat{f}(  k)\vert ^{2} d  k $,  the wavelet energy spectrum, i.e.  the scale distribution of the emission
is defined as the energy in all wavelet
coefficients of scale {\it a},
\begin{equation}
M(a) = \int_{-\infty}^{+\infty} \int_{-\infty}^{+\infty}  \vert W(a,{  x})\vert ^{2} d  x,
\end{equation}
\noindent which is expressed in Fourier space as
\begin{equation}
M(a) = \frac{a^{4-2\kappa}}{16\pi^4} \int_{-\infty}^{+\infty} \int_{-\infty}^{+\infty} \vert \hat{f}(  k)\vert ^{2} \vert \hat{\psi}(a   k) \vert^2  d  k.
\end{equation}

We use  $\kappa=2$, which provides the same power law for the wavelet spectrum as for the conventional second-order structure function \citep[][]{Frick_etal_01}. This selection also ensures a consistent comparison with previous studies \citep[e.g.][]{Hughes_etal_06,Tabatabaei_1_07,Dumas}.

By decomposing the images into maps containing the structures of a
given scale we can analyze the cross-correlation of the analyzed maps
scale-by-scale \citep{nesme,Frick_etal_01}. The wavelet
cross-correlation coefficient at scale {\it a} is defined for 2D
maps as
\begin{equation}
r_{w}(a)=\frac{\int \int W_{1}(a,{  x})~W^{\ast}_{2}(a,{  x}) d{  x}}{[M_{1}(a)M_{2}(a)]^{1/2}},
\end{equation}
\noindent where the subscripts refer to two images of the same size and same
resolution. The value of $r_{w}$  varies between $-1$ (perfect
anticorrelation) and $+1$ (perfect correlation). We will consider
$|r_{w}|=0.5$ as a marginal value for the acceptance of a
correlation between the structures of given scale\footnote{Generally, $|r_{w}|=0.5$ indicates the same number of
correlated and uncorrelated structures in two decomposed maps. }.
Plotting $r_{w}$ against scale shows how well structures at
different scales are correlated in intensity and location.

The error of an estimated correlation depends on the number of
independent points used, $n$, and on the degree of correlation
\citep{Edwards}
\begin{equation}
\Delta r_{w}(a)= \frac{\sqrt{1-r^{2}_{w}}}{\sqrt{n-2}},
\end{equation}
where the number of independent points is estimated as $
n\,\simeq\,L_x L_y / a^2$ and $(L_x, L_y)$ are the map sizes in
$(x,y)$. Thus, towards larger scales, $n$ decreases and the errors
increase. For example, considering the size of our radio and IR maps
for M~33, less than 5 independent points exist for scales
$a>20\arcmin$. After calculating the correlations, the t-student
value $t=r_w\,\sqrt{(n-2)/(1-r_w^2)}$ \citep[e.g.
][]{Hoernes_etal_98} is smaller than 2 for $a>20\arcmin$,  which
means that the correlation is not statistically significant.
Therefore, we only consider wavelet correlations for scales smaller
than $20\arcmin$.
\begin{figure}
\begin{center}
\resizebox{7cm}{!}{\includegraphics*{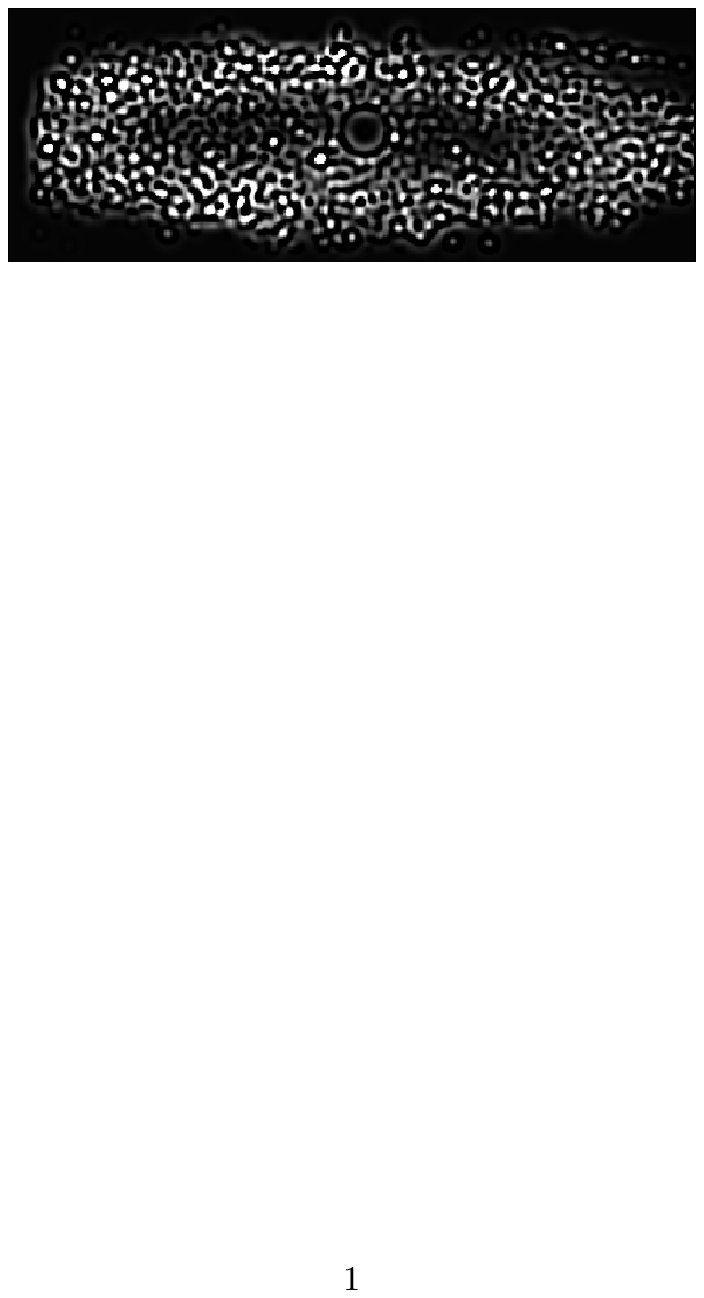}}
\resizebox{7cm}{!}{\includegraphics*{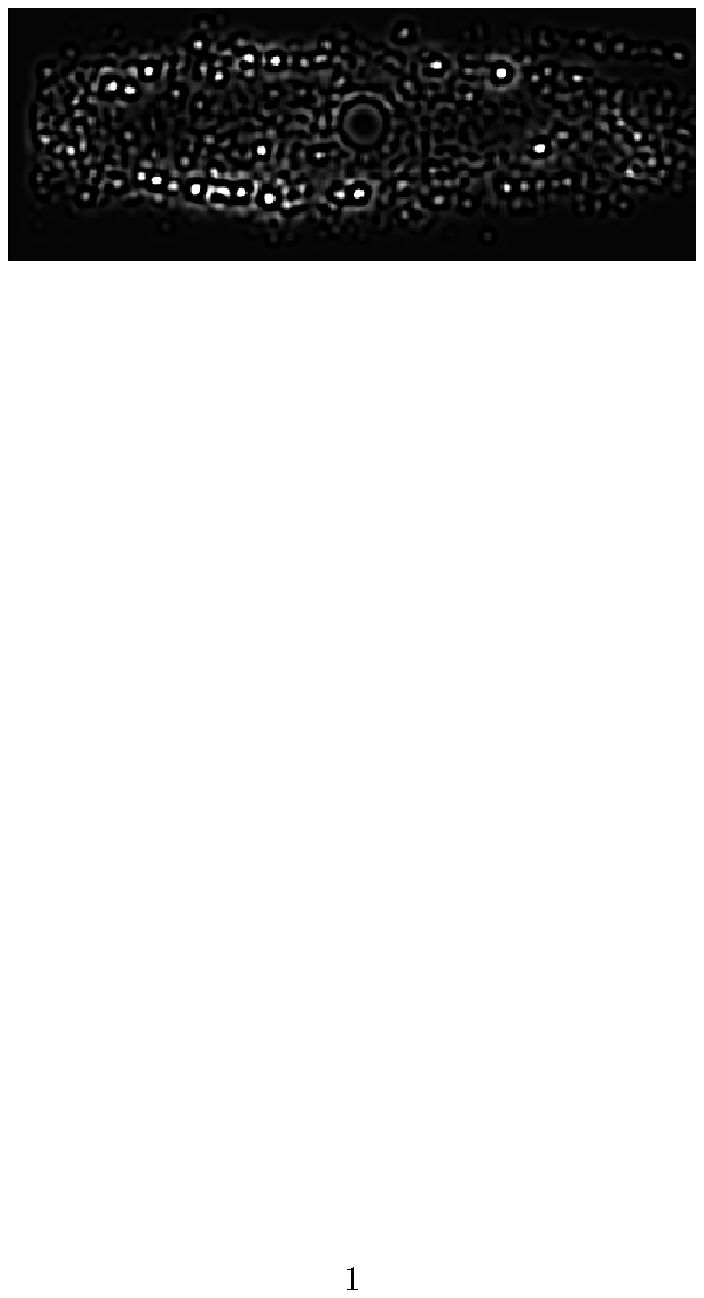}}
\resizebox{7cm}{!}{\includegraphics*{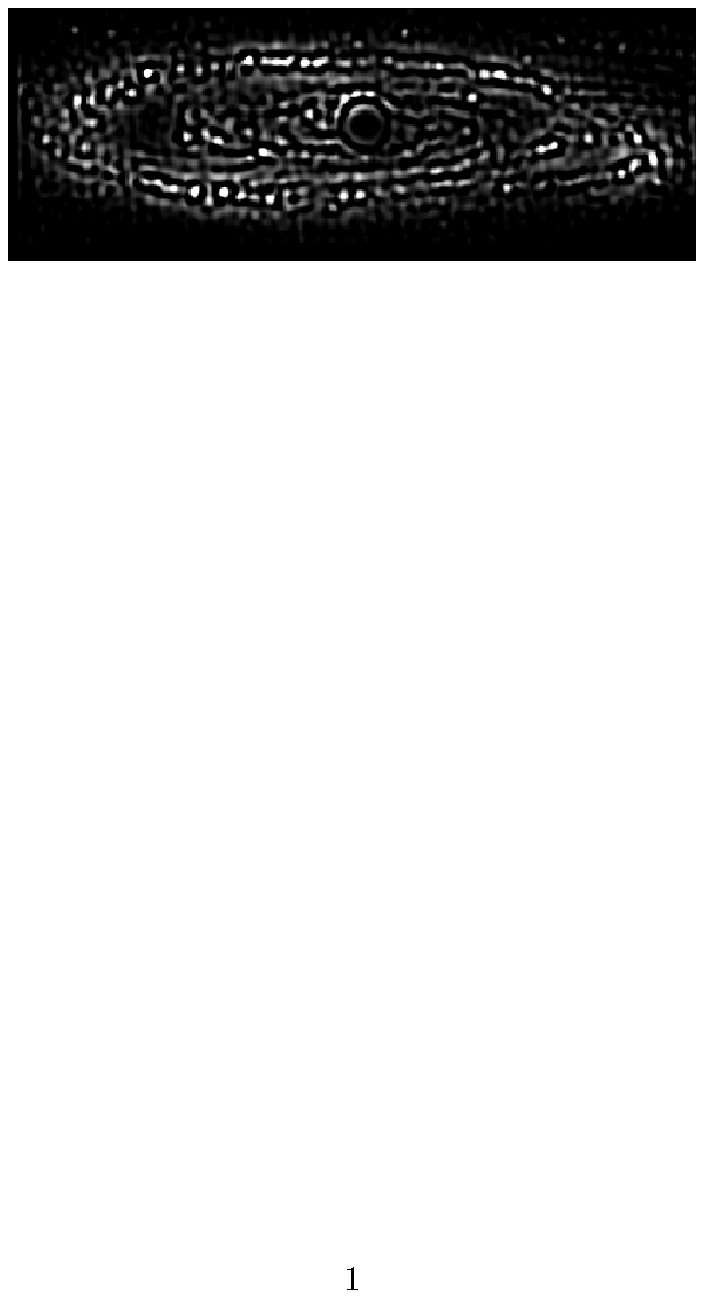}}
\resizebox{6.95cm}{!}{\includegraphics*{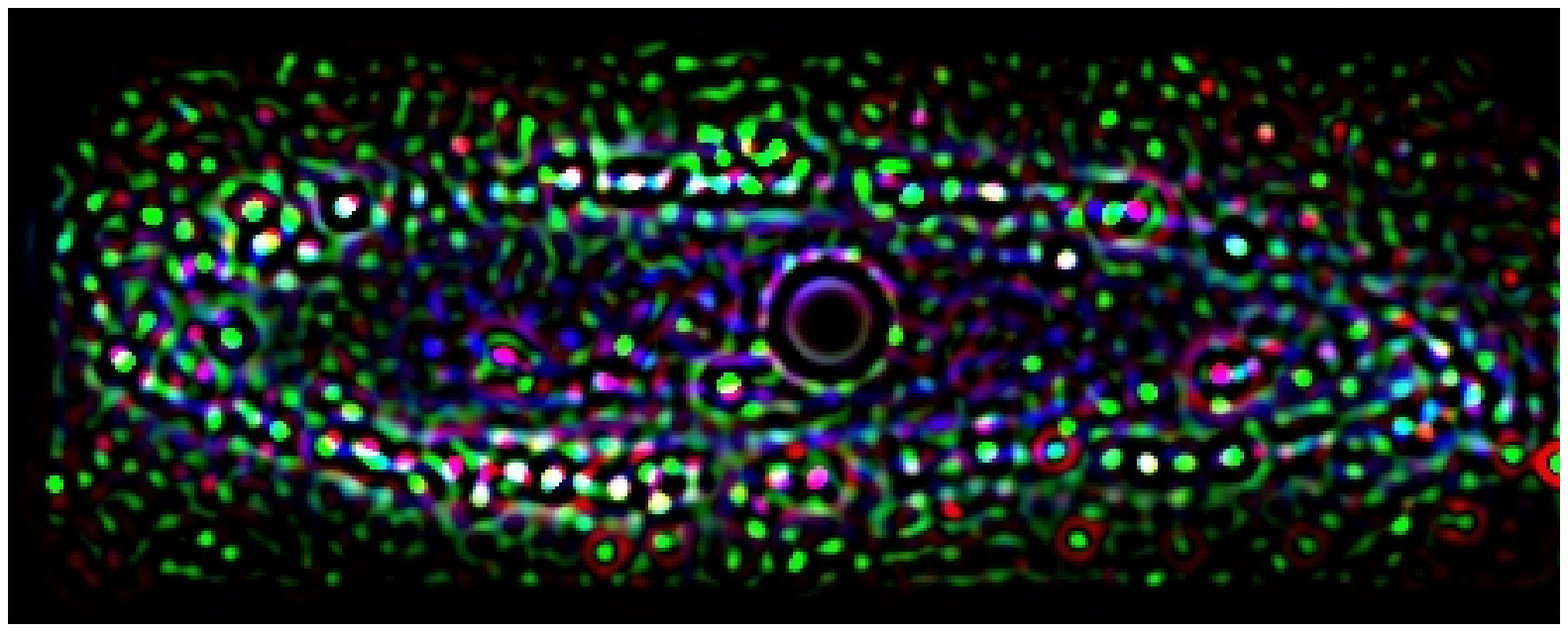}}
\caption[]{M~31 maps of the synchrotron
(20\,cm, {\it top}), free-free (20\,cm, {\it middle}), and
160\,$\mu$m ({\it bottom}) emission on the smallest scale of the
wavelet decomposition, $a=1.7\arcmin \simeq\,0.4$\,kpc {  in arbitrary units}. The emission from the central 2\,kpc was removed before decomposition.  The map size is
$110\arcmin \times 38.5\arcmin$. {  Also shown is the composite map (red: free-free, green: synchrotron, and blue: 160\,$\mu$m emission)}. }
\label{fig:decompose}
\end{center}
\end{figure}
\begin{figure}
\begin{center}
\resizebox{6.75cm}{!}{\includegraphics*{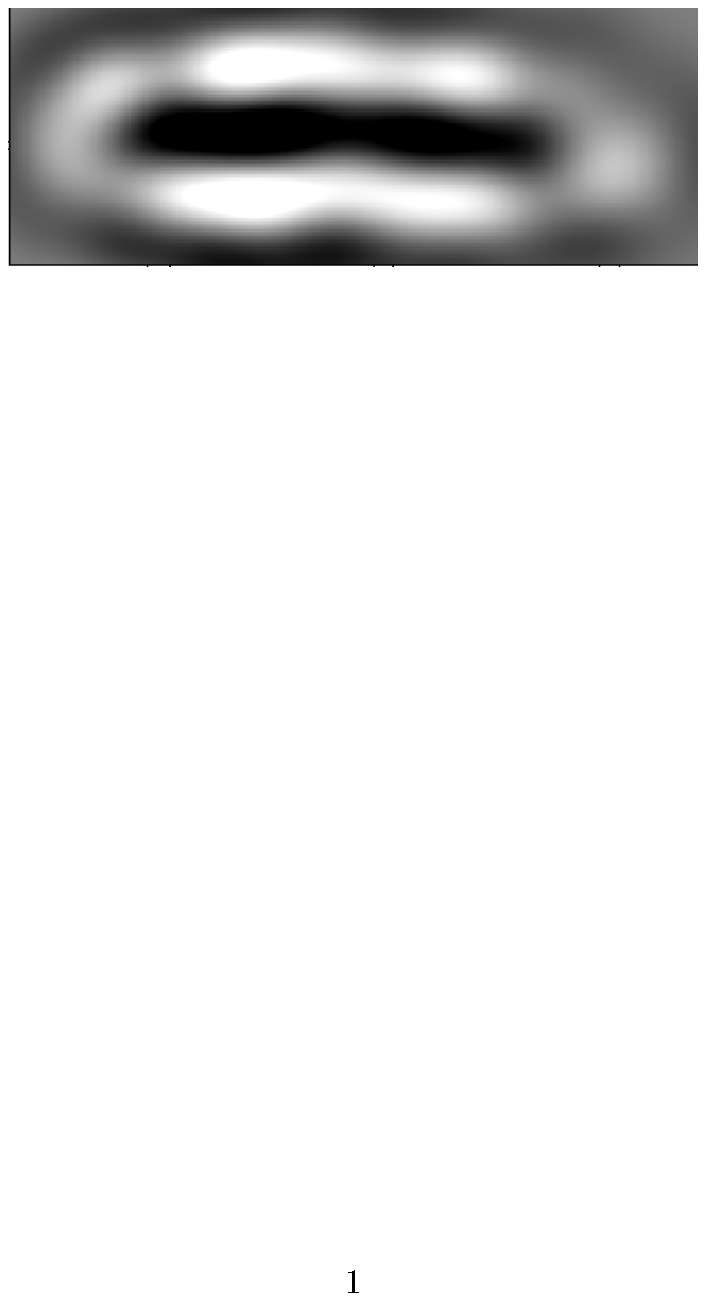}}
\resizebox{6.75cm}{!}{\includegraphics*{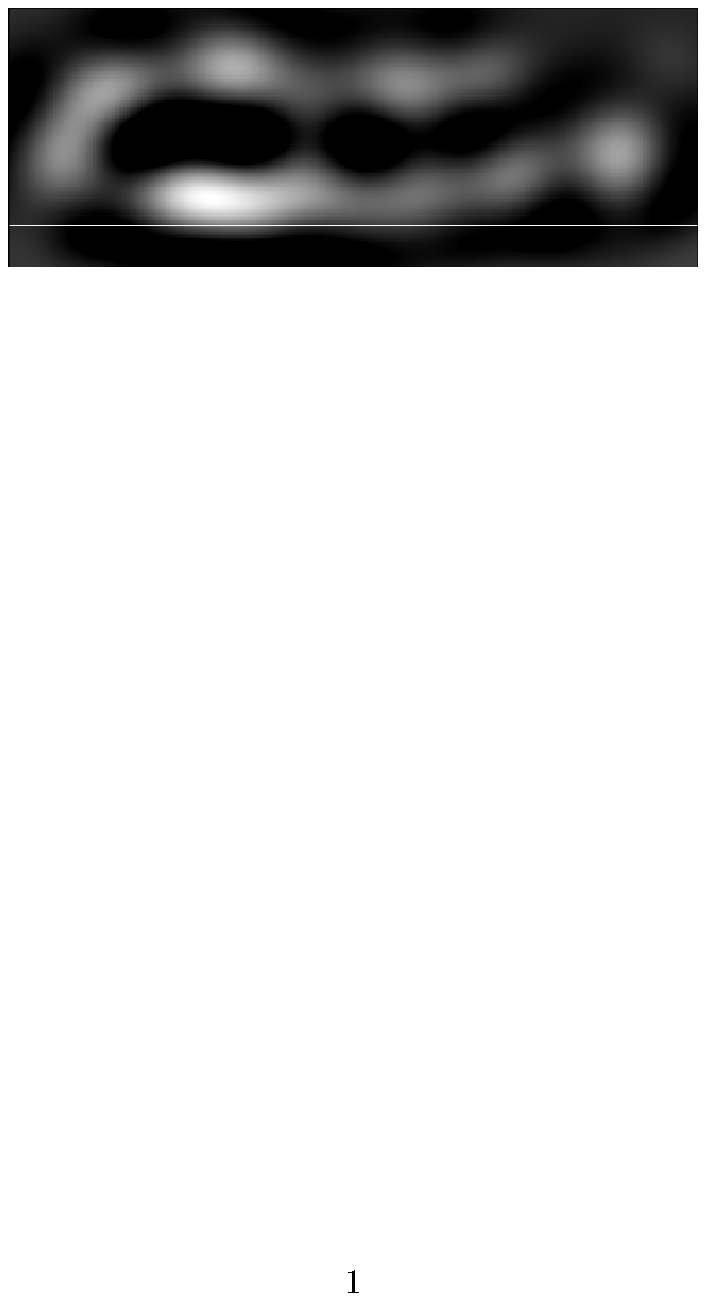}}
\resizebox{6.75cm}{!}{\includegraphics*{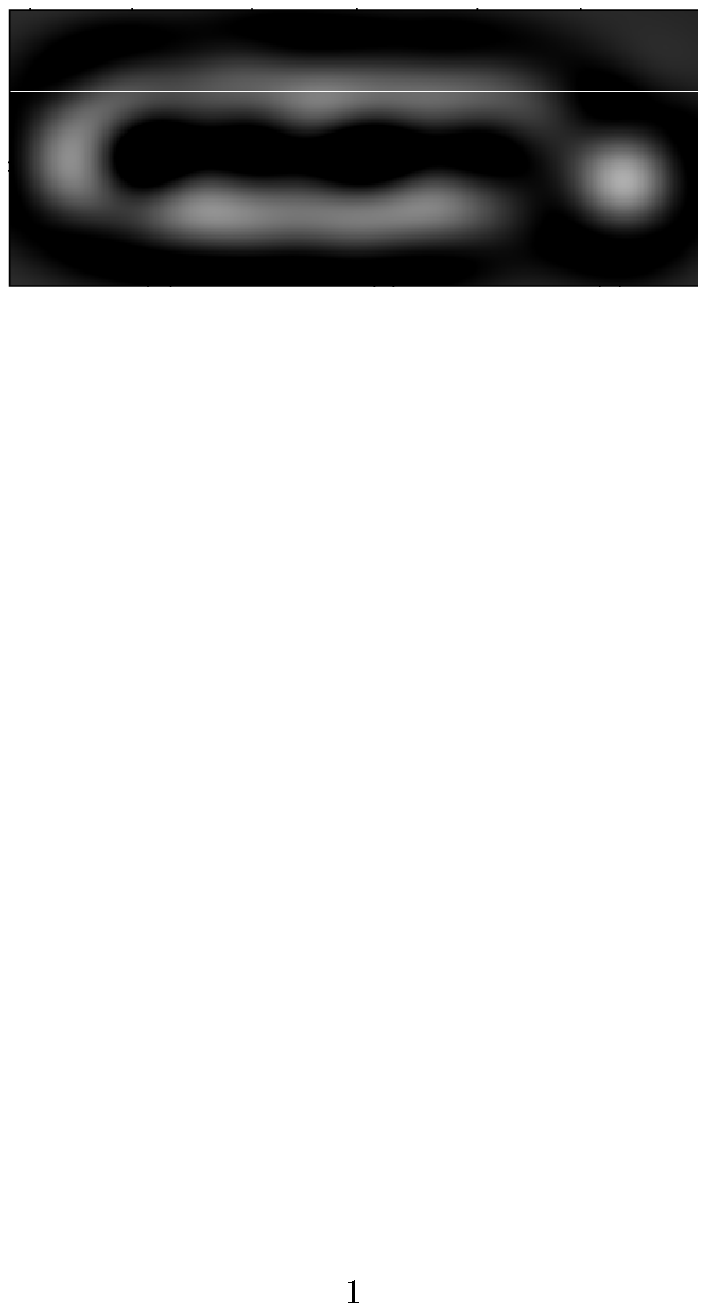}}
\resizebox{6.75cm}{!}{\includegraphics*{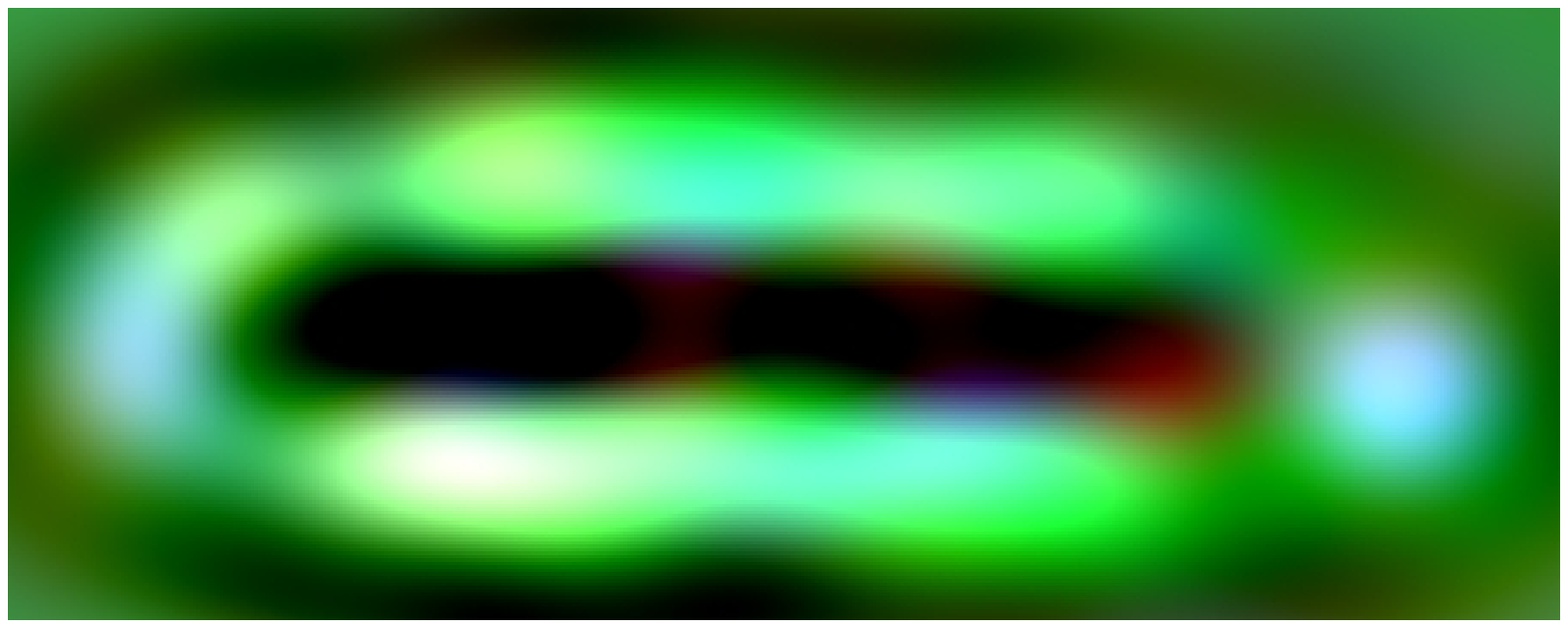}}
\caption[]{The same as Fig. 5 for a large scale of $a=11\arcmin\simeq\,2.5$\,kpc.  }
\label{fig:decompose}
\end{center}
\end{figure}

\begin{figure*}
\begin{center}
\resizebox{10.5cm}{!}{
\includegraphics*{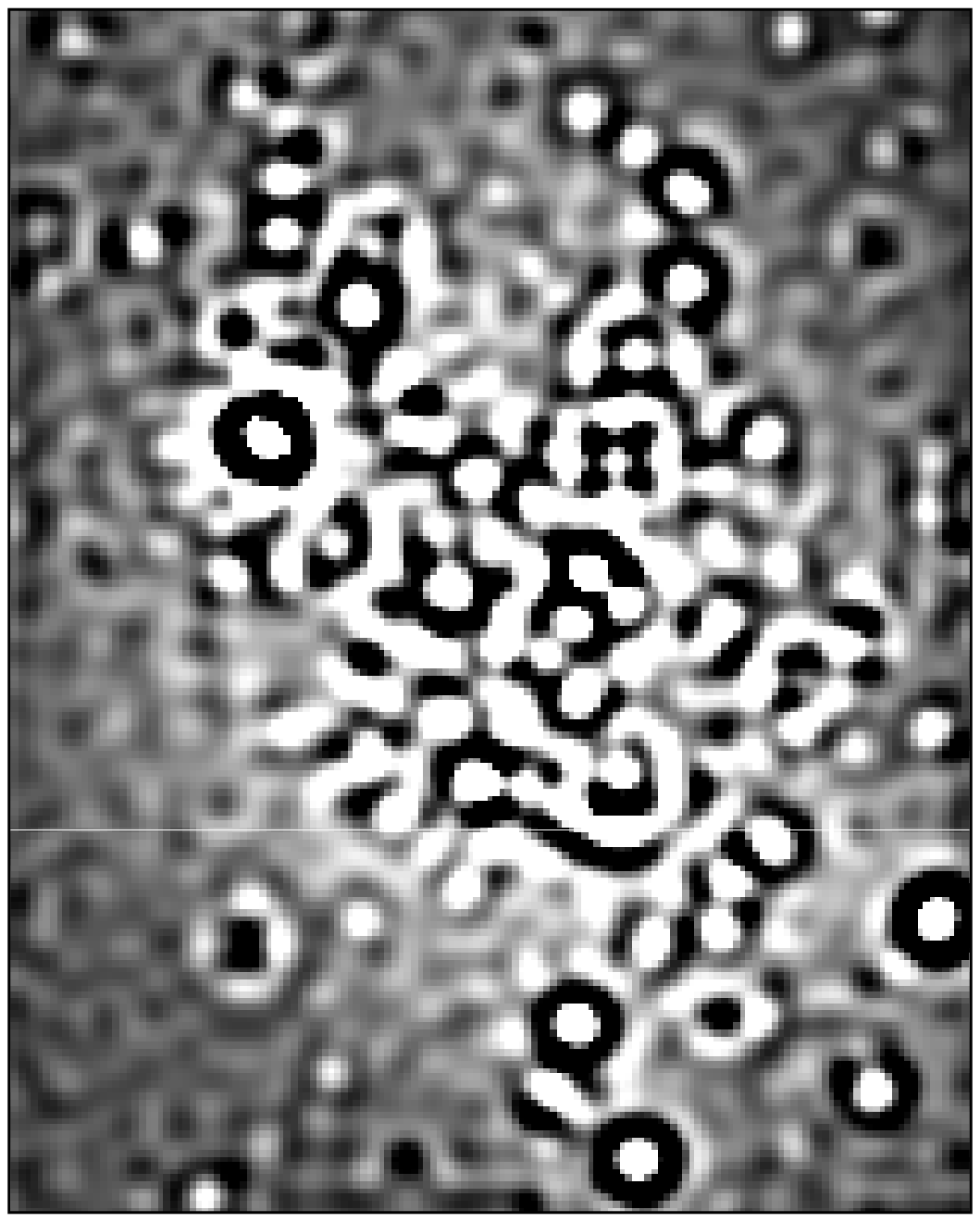}
\includegraphics*{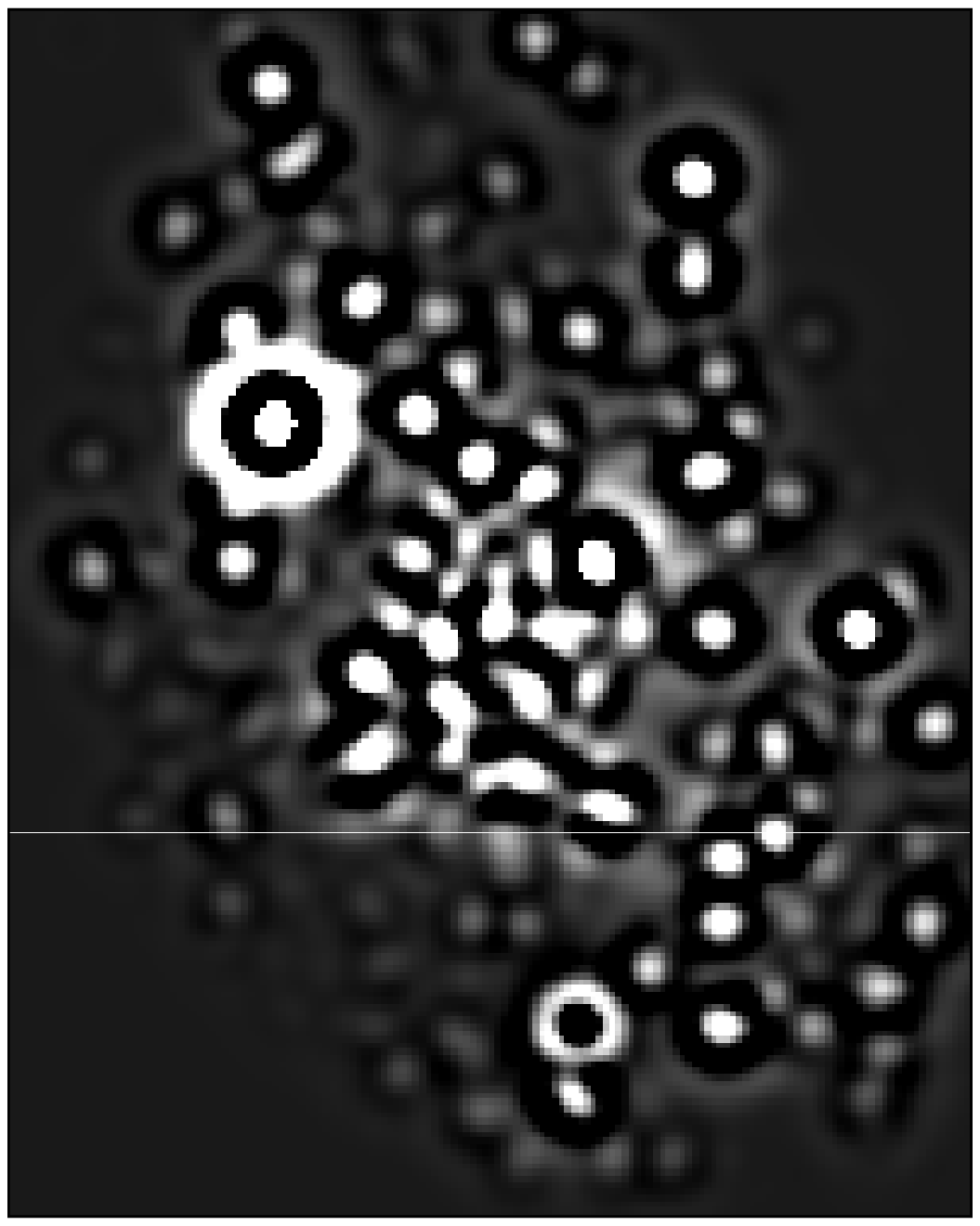}
\includegraphics*{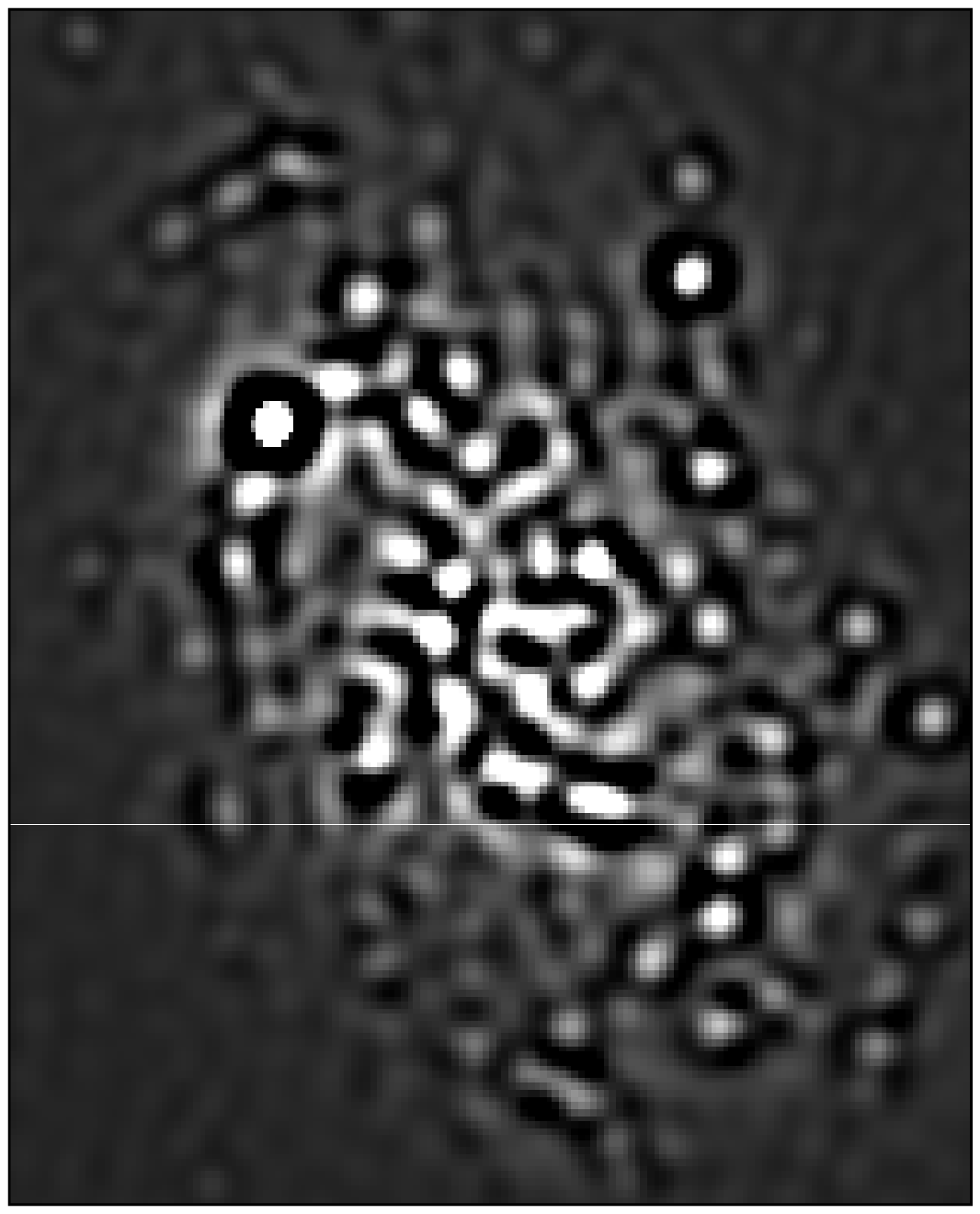}}
\resizebox{3.48cm}{!}{\includegraphics*{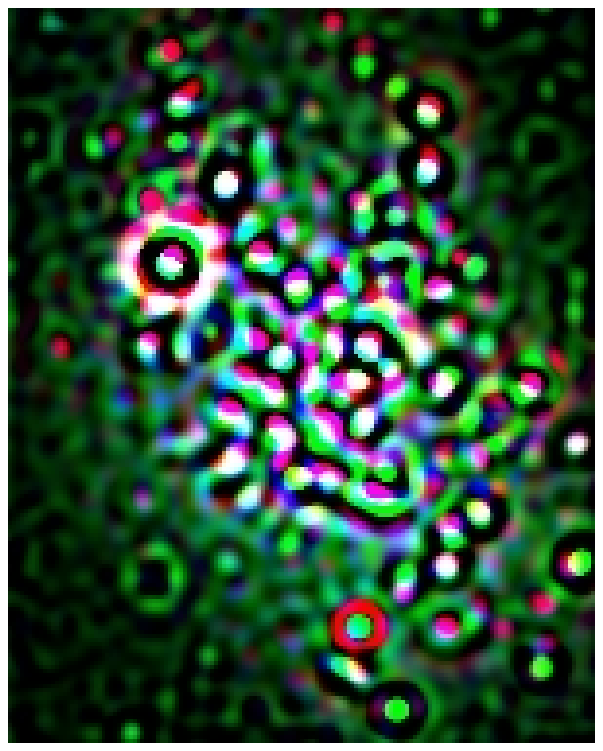}}
\resizebox{10.5cm}{!}{
\includegraphics*{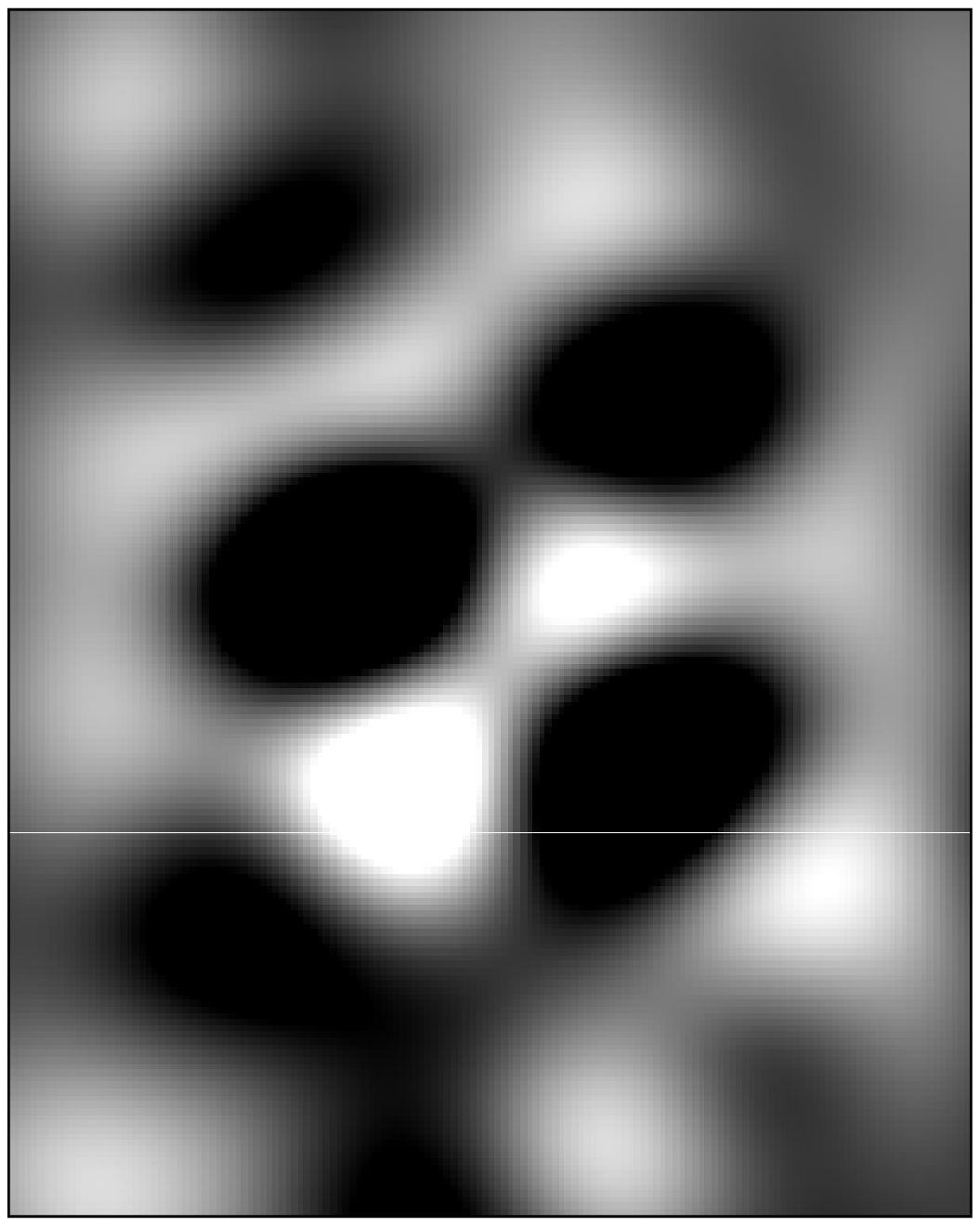}
\includegraphics*{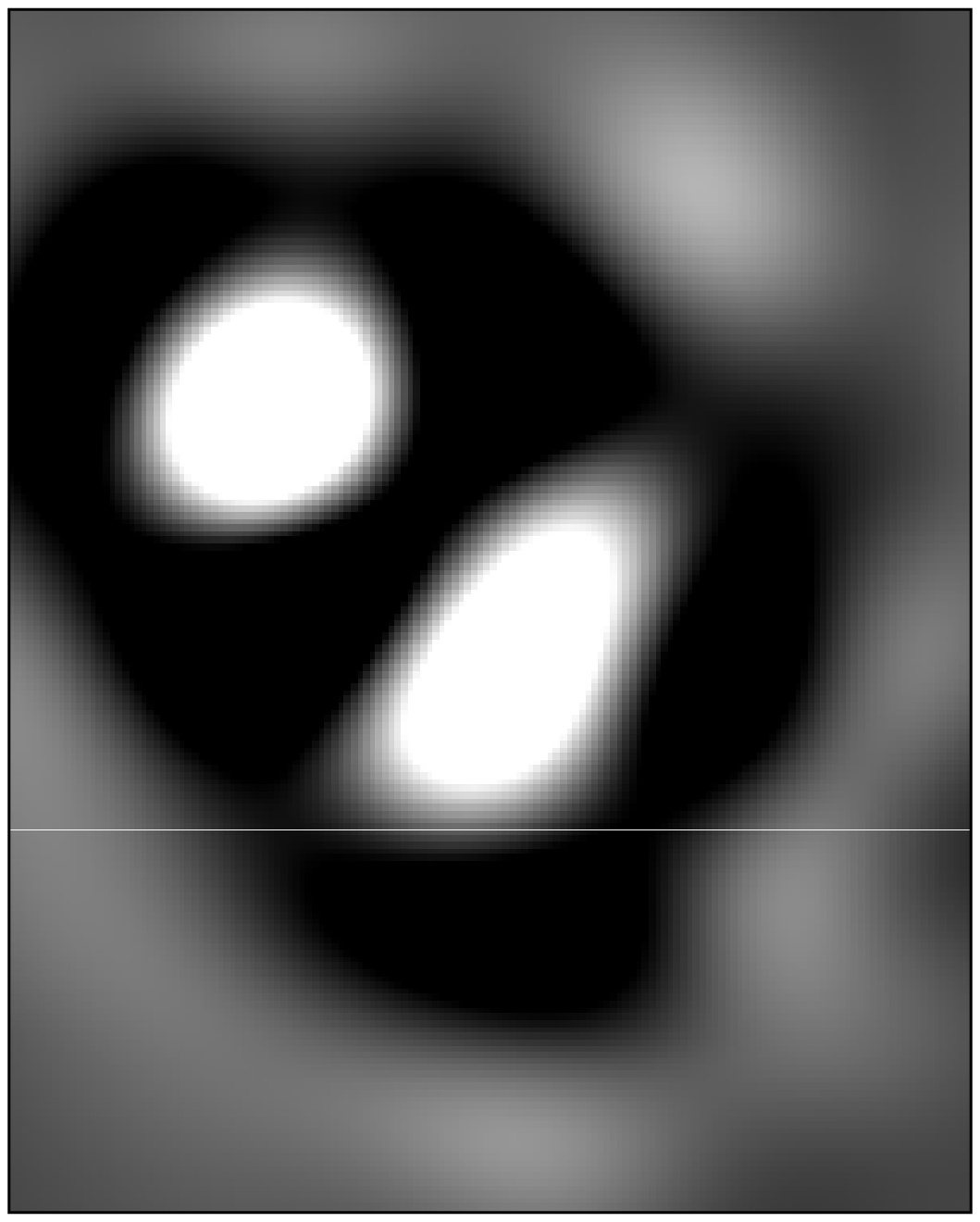}
\includegraphics*{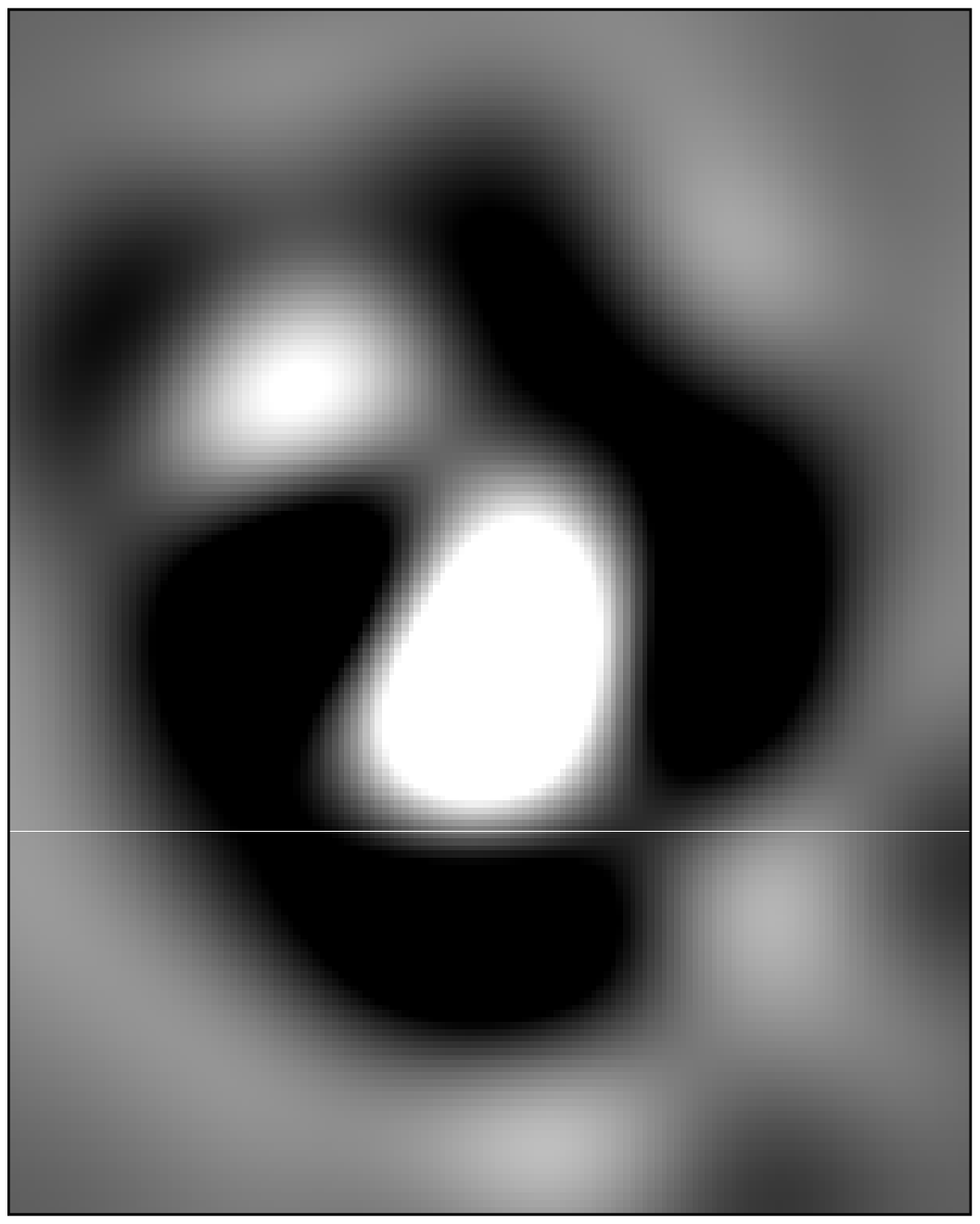}}
\resizebox{3.5cm}{!}{\includegraphics*{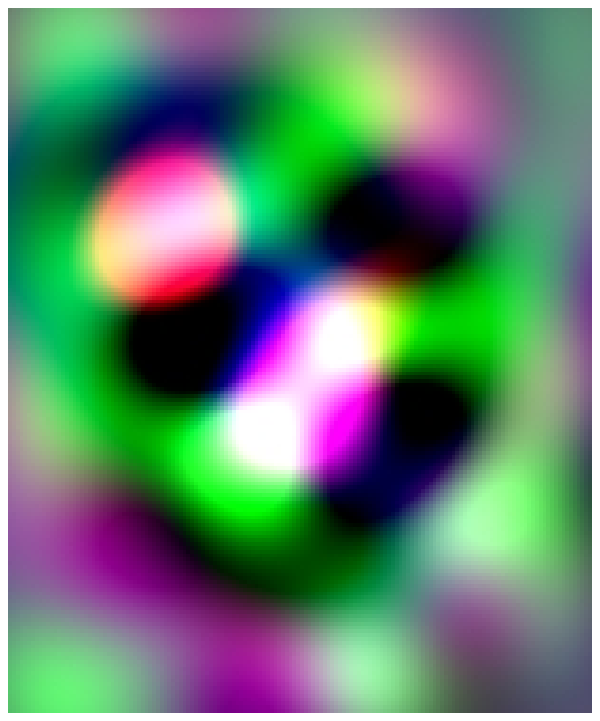}}
\caption[]{
M~33 maps of the synchrotron (20\,cm, {\it left}), free-free
(20\,cm, {\it middle}), and 160\,$\mu$m ({\it right}) emission
on the  smallest scale of $a
\sim\,0.4$\,kpc ({\it upper row}) and a large scale of $a=12.5\arcmin \simeq\,3$\,kpc ({\it bottom row}) {  in arbitrary units}. The map size is
$42\arcmin$ x $54\arcmin$. {  Also shown are the composite maps (red: free-free, green: synchrotron, and blue: 160\,$\mu$m emission)}. }
\label{fig:decompose2}
\end{center}
\end{figure*}

To illustrate what kind of information can be deduced from wavelet
cross-correlation analysis we return to our test example and show
in Fig.~\ref{fig:test} the wavelet spectrum of the test image and
the wavelet cross-correlation function. The spectrum shows two
maxima corresponding to the scale of small spots and to the scale
of a cloud of spots. In general, the analysis of the
correlations $r_w(a)$ only makes sense for energy-containing scales,
i.e. in the vicinity of the two peaks in the example
discussed.

Actually, $r_w\approx -1$ near the first peak (the scale of small
spots, (their locations differ in both maps) and $r_w \approx 1$
near the second peak (the {  location of the cloud of spots in} both maps is the
same).

Note that by using a continuous wavelet transform we formally
can consider a continuous range of scales (or a set of scales with
arbitrary small steps). However, the scale resolution of the wavelet
is finite, which means that the adjacent scales are not independent and it
is not correct to take a very dense sampling in scales. Actually,
the wavelet (Eq.~\ref{pethat}) used below covers an octave in
Fourier space, which means that a reasonable sampling corresponds to
a scale doubling at each step. The samples used below are close to
this {  criterion}.
\subsection{Wavelet spectra and cross-correlations for M~31 and M~33}
\begin{figure*}
\begin{center}
\resizebox{14cm}{!}
{\includegraphics*{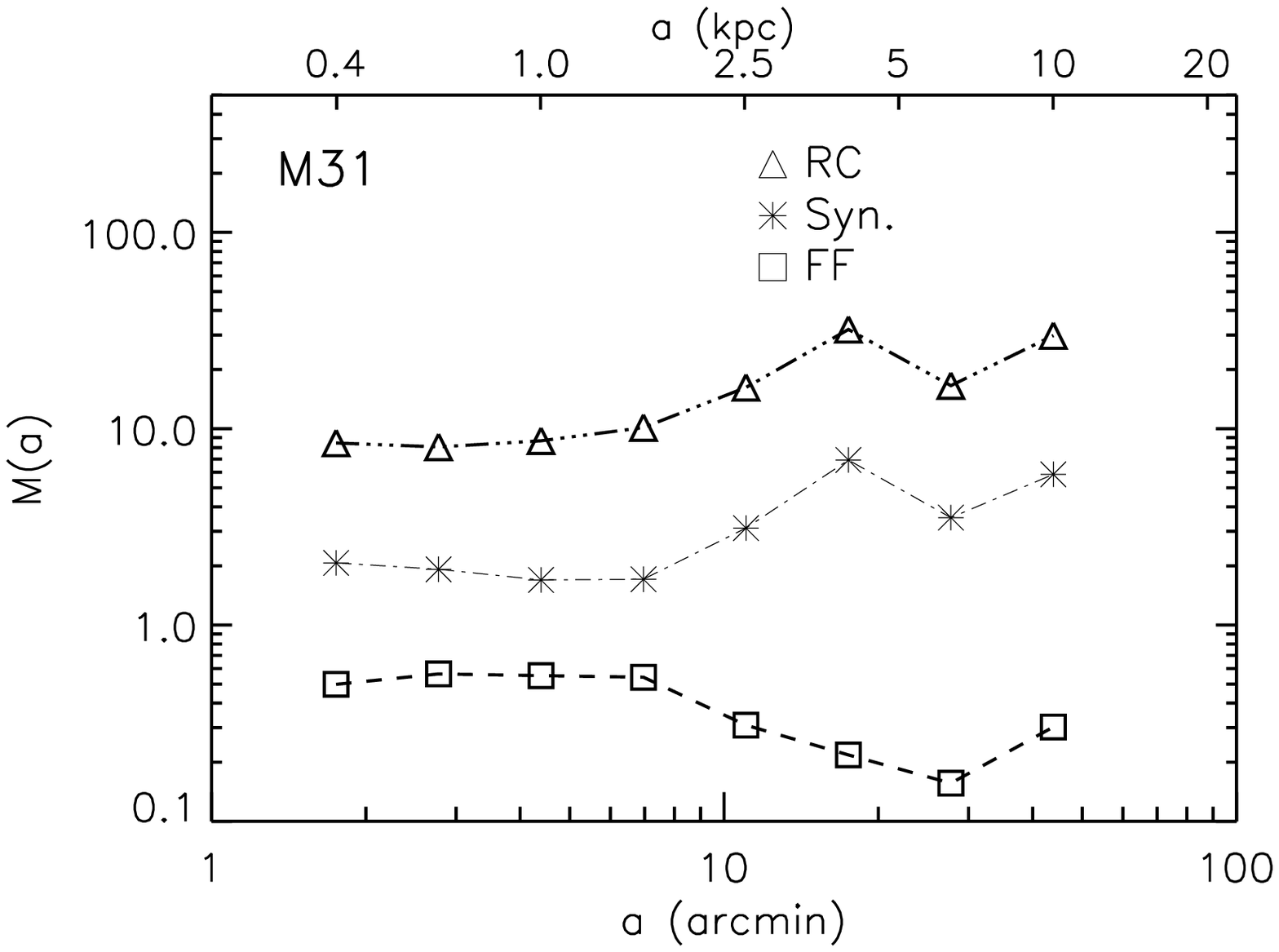}
\includegraphics*{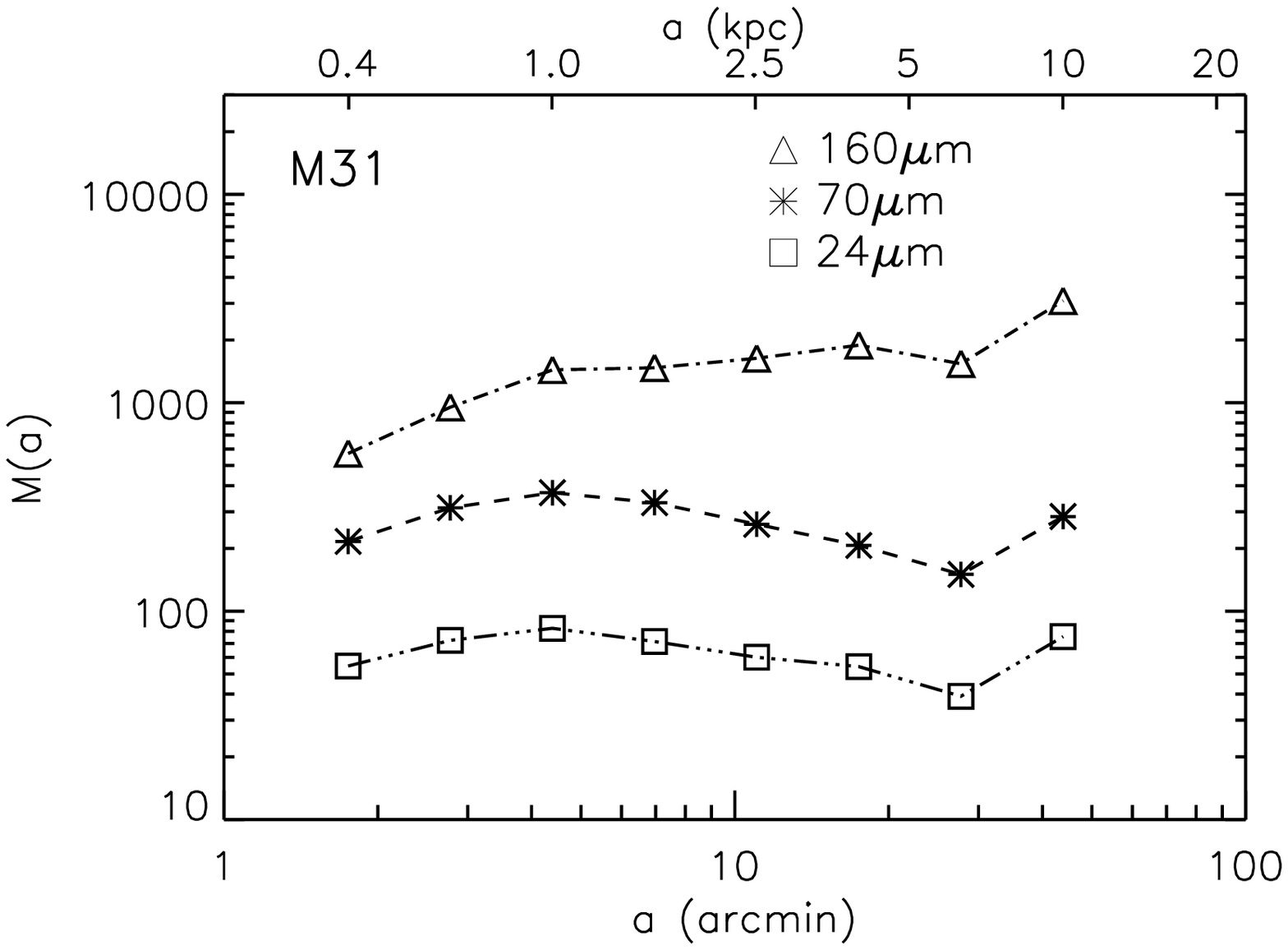}}
\resizebox{14cm}{!}
{\includegraphics*{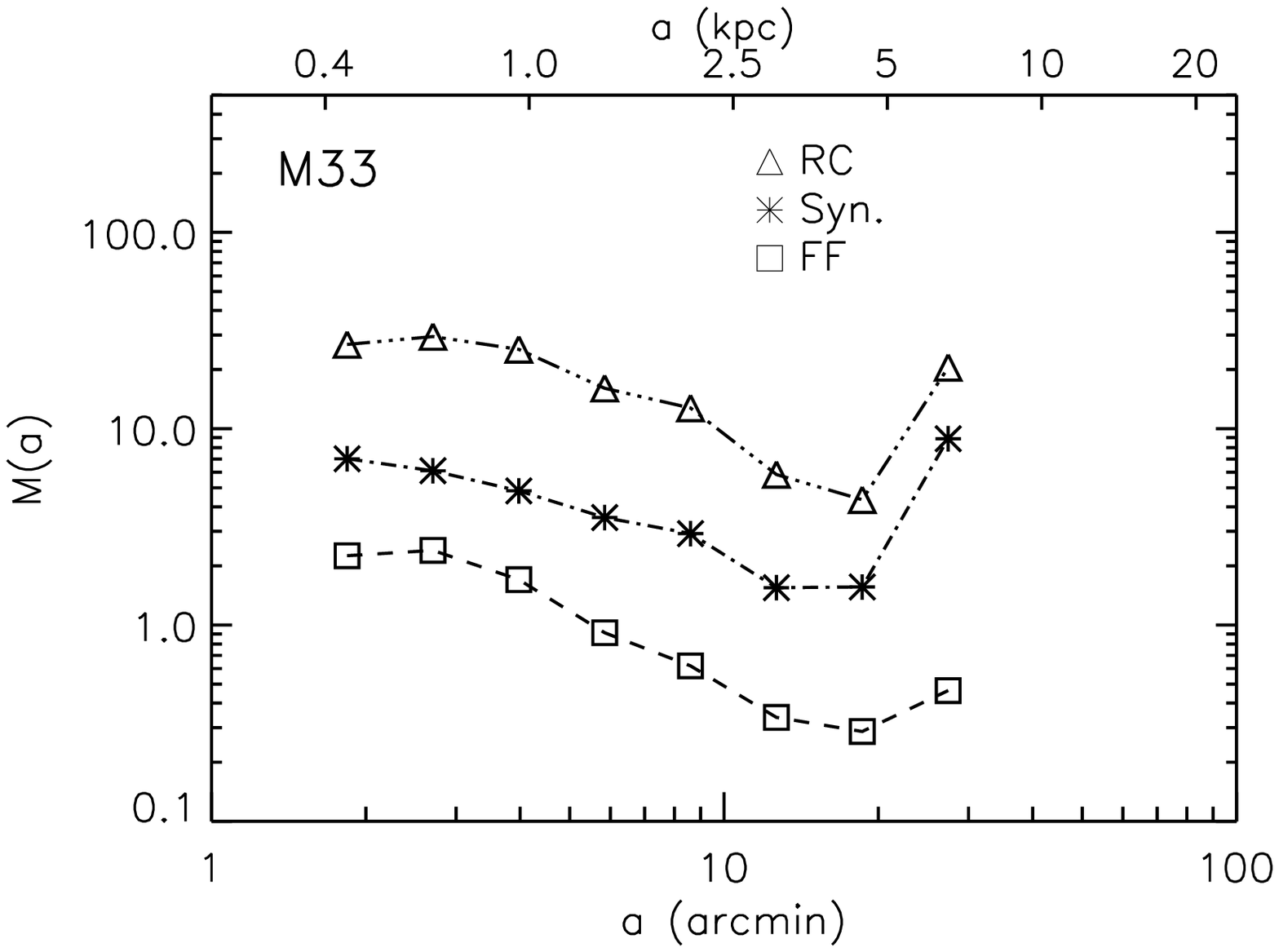}
\includegraphics*{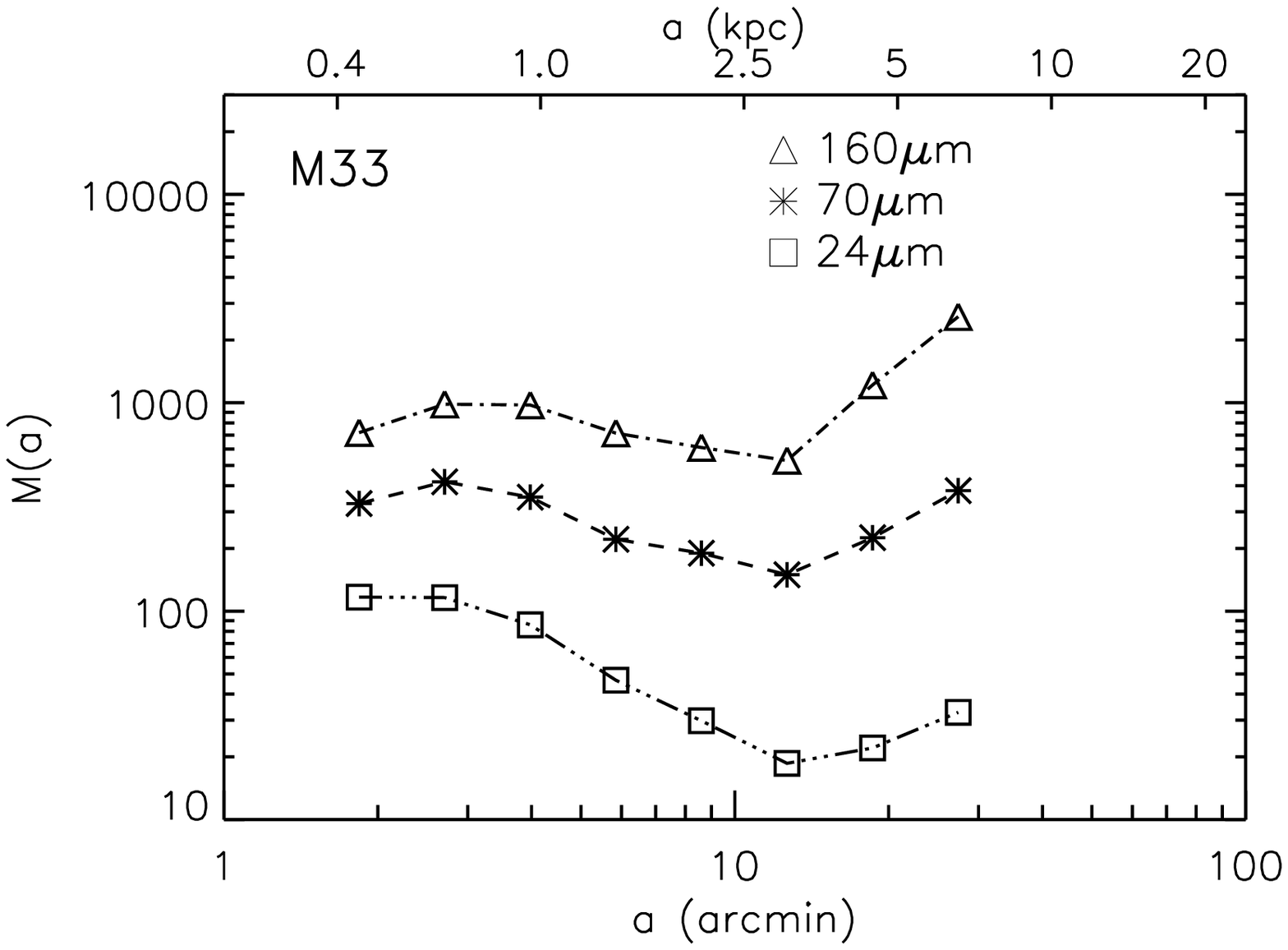}}
\caption[]{Wavelet spectra of the observed 20\,cm radio continuum
(RC), synchrotron (Syn.), and free-free (FF) emissions ({\it left})
and IR emissions ({\it right}) from M~31 ({\it top}) and M~33
({\it bottom}), shown in arbitrary units. The errors calculated by adding
Gaussian noise (with an amplitude of one $\sigma$ noise level of the
maps) before the decomposition are smaller than the symbols. }
\label{fig:wave1}
\end{center}
\end{figure*}

We decomposed the 20\,cm radio and the MIPS IR maps of the two
galaxies into 8 angular scales ({\it a}). Before decomposition, the
central region (R$<1$\,kpc) in the M~31 maps was subtracted to avoid
a strong influence of the nucleus on the results.

The decomposed maps of the synchrotron, free-free, and 160\,$\mu$m emissions are shown in Figs.~5 and 6 for M~31 and Fig.~7 for M~33, on the smallest scale of $1.7\arcmin\ \simeq\,0.4$\,kpc (about twice
the beamwidth of the observed maps) and on a selected large scale of $a=12.5\arcmin\
$($\simeq$\,3\,kpc) {  for M~33 and $a=11\arcmin\ $($\simeq$\,2.5\,kpc) for M~31}.
On the smallest scale, resolved sources (e.g. HII regions) appear as distinct point-like
structures, but narrow features of diffuse emission are also
visible. On $a \simeq$\,3\,kpc, only bright large-scale (extended) structures are exhibited.

In M~31, a wide distribution of small-scale structures (see Fig.~5) occurs in the synchrotron
map at the position of the `10\,kpc ring',
while the free-free emission is dominated by `points' coinciding
with SF regions (they are at the same positions as the SF regions in Fig.~2, top). At 160\,$\mu$m,
the small-scale structures,   are aligned along the spiral arms and are bright in the SF regions {  (see also the composite image in Fig.~5)}.
The fact that the small-scale synchrotron emission does not follow the narrow arms
indicates that CREs propagated away from their origin (see also Fig.~2, bottom).
On the large scale of $\simeq$\,3\,kpc (Fig.~6), the  wide
`10\,kpc ring' is now dominant in all emissions.

In M\,33,  the small-scale synchrotron emission is
widely distributed across the entire disk, yet it is bright near the SF
regions that are well visible in the free-free and the 160\,$\mu$m decomposed maps (Fig.~\ref{fig:decompose2}, upper row).
On the large scale of $\simeq$\,3\,kpc, the two bright structures\footnote{  Their adjacent
dark regions are due to the negative part of the wavelet function applied.}  in the free-free and 160\,$\mu$m emission, seen near the center and on the position of the HII region NGC\,604,
are not visible in the synchrotron emission (Fig.~\ref{fig:decompose2}, bottom row).

Hence, the morphological difference between the FIR 160\,$\mu$m and the synchrotron emission is
more evident on the small scale of 0.4\,kpc  than on the large scale of 3\,kpc in M~31, while it
is the other way around in M~33. Generally,  the FIR emission has a morphology  more similar to
that of the free-free than of the synchrotron emission in both galaxies.

To investigate which spatial scales dominate the emissions, we plot
the wavelet spectra M(a) versus spatial scale \footnote{Since M(a)
is a relative measure, standard deviations cannot be computed. The
systematic effect of noise on small scales \citep{Dumas}, probed
by adding Gaussian noise to the maps before decomposition, is
negligible for M~31 and M~33 on scales shown in
Fig.~\ref{fig:wave1}.} for the MIPS IR bands and the 20\,cm radio
emission in Fig.~\ref{fig:wave1}. In general, the presence of strong
point-like sources causes a maximum on small scales. In case of a
dominant component of extended diffuse emission, the spectrum
increases towards larger scales.

In M~31 (Fig.~\ref{fig:wave1}, upper panels), the observed RC
emission peaks at $a\simeq$\,4\,kpc, which is the width of the
bright synchrotron ring at $R\simeq 10$\,kpc. For the
synchrotron emission small scales are less important than the larger
scales,  unlike for the free-free emission. The free-free emission
is dominated by sources on  $a\,\le 2$\,kpc, residing mainly in
M~31's `10\,kpc ring'. Among the IR bands, the 70\,$\mu$m spectrum
is most similar to that of the free-free emission.
The 160\,$\mu$m
emission is entirely dominated by large scales indicating the
importance of the ISRF in heating the cold dust.

M~33 shows a different behavior. The wavelet spectra of the
24\,$\mu$m and the free-free emission are most similar
(Fig.~\ref{fig:wave1}, lower panels). Here even the cold dust emits
significantly on small scales. Moreover, the radio emission, and
even its synchrotron component, {  is}  relatively important on small
scales. Comparing the (Fourier) spectral energy densities in the band
$0.4<a<2$\,kpc in the two galaxies, we find that the synchrotron emission on these scales
is a factor 1.7--2 larger in M~33 than in M~31. This shows that the magnetic field
strength and/or the CRE density are higher on small scales in M~33.
Due to the high SF activity in M~33,  synchrotron, free-free and
warm-dust emission are significant on small scales and show a fast
decrease towards $a\simeq3-4$\,kpc, after which an increase on
larger scales occurs.

\begin{figure}
\begin{center}
\resizebox{7cm}{!}
{\includegraphics*{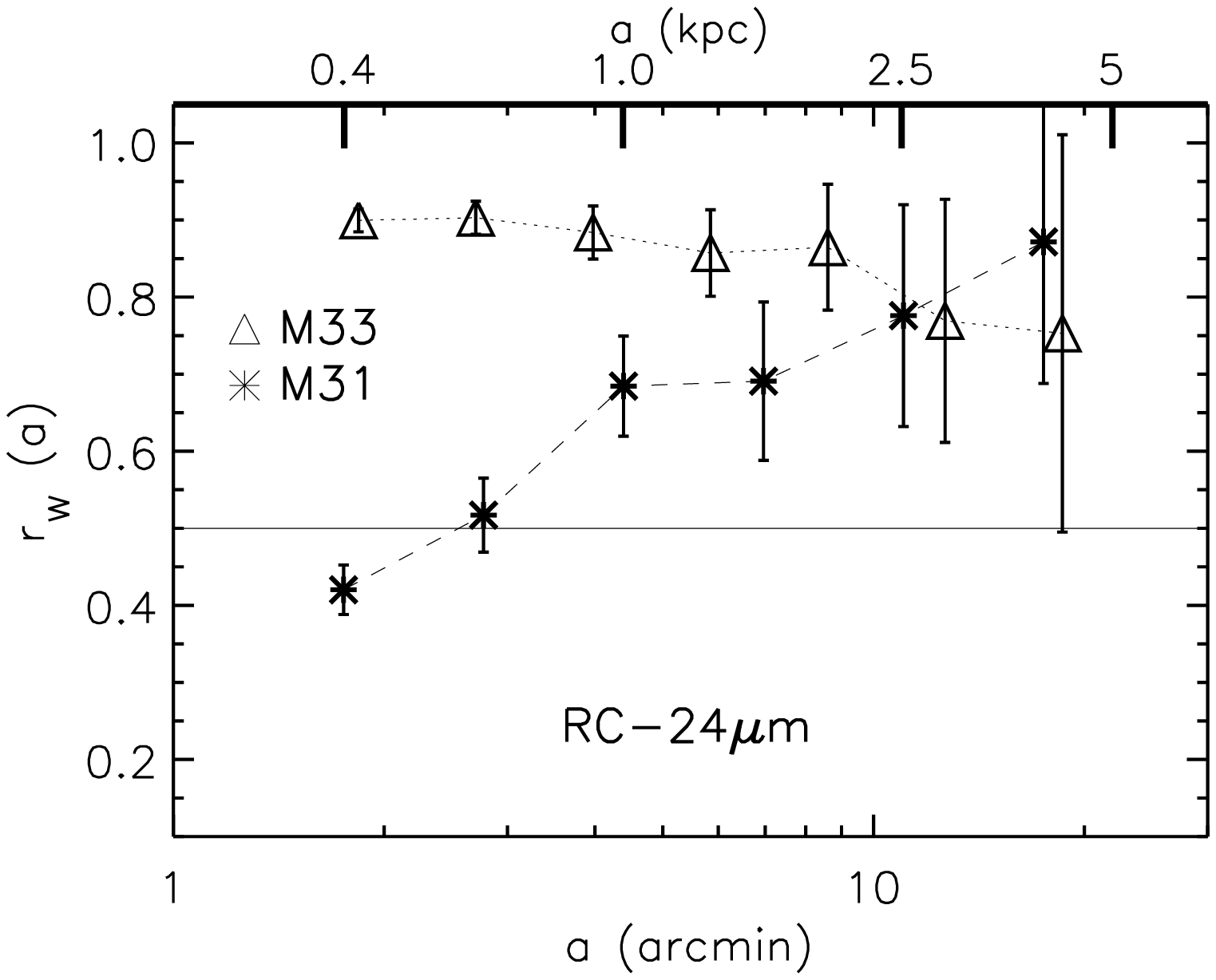}}
\resizebox{7cm}{!}
{\includegraphics*{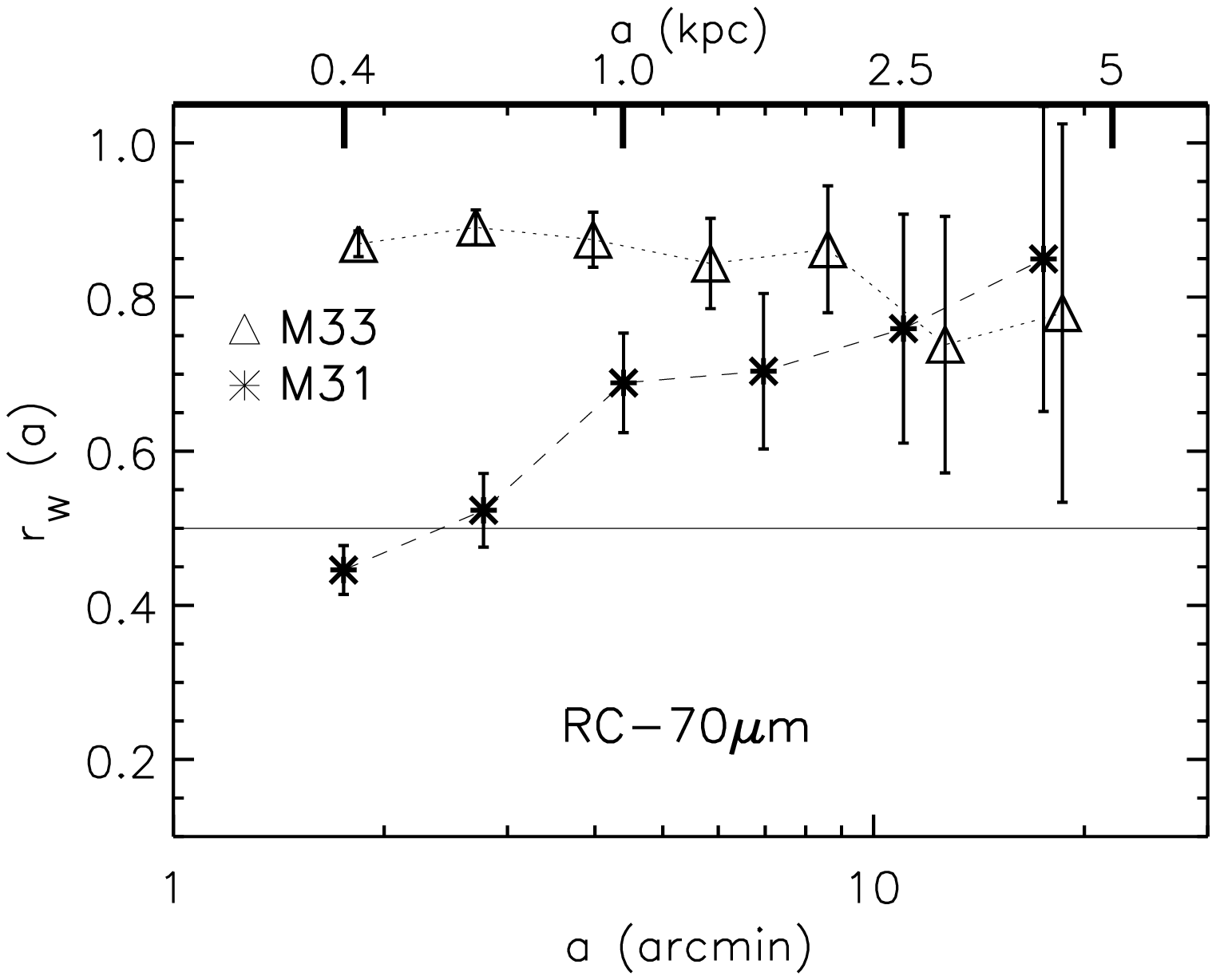}}
\resizebox{7cm}{!}
{\includegraphics*{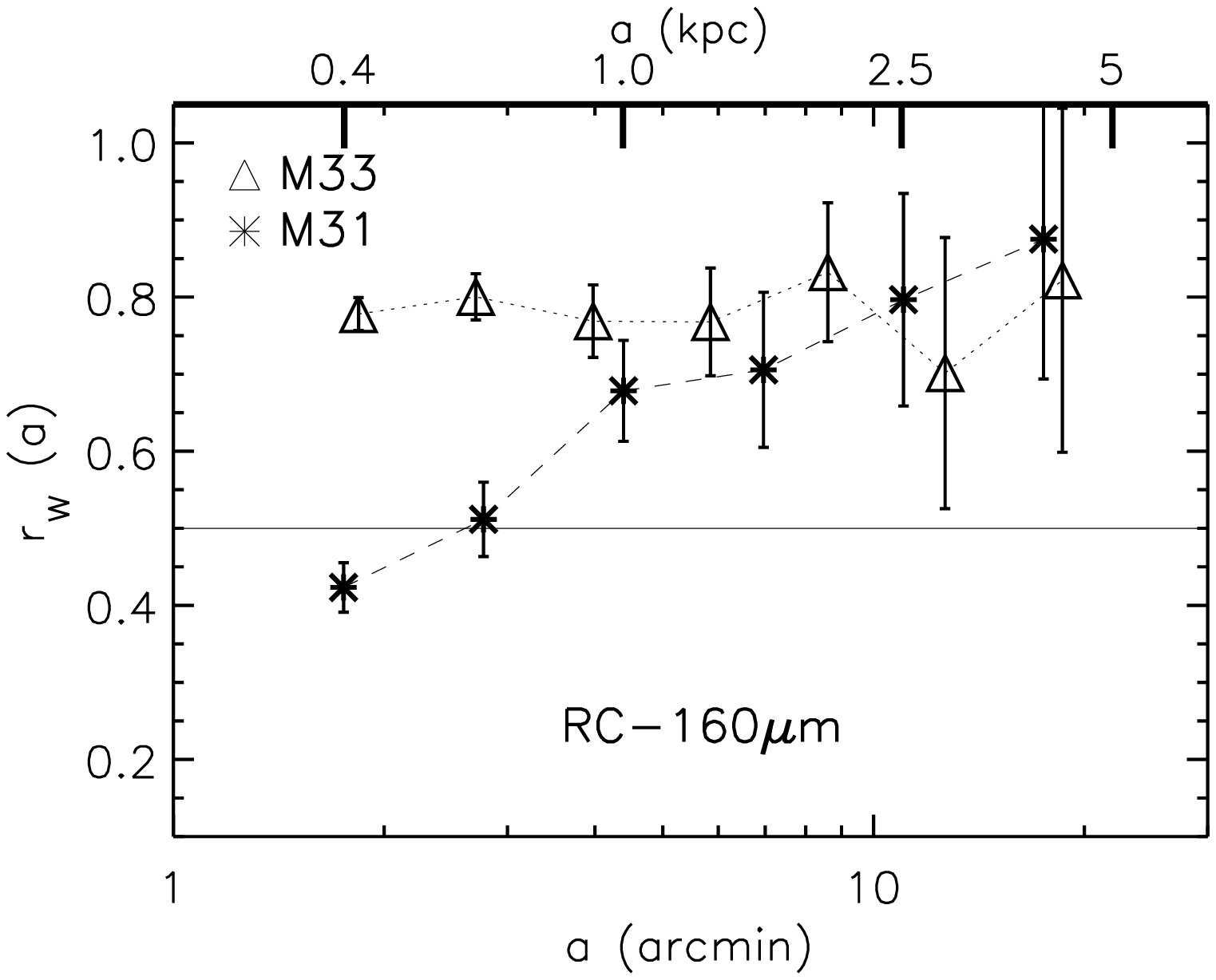}}
\caption[]{Scale-by-scale cross-correlations between IR and observed
20\,cm radio continuum (RC) emission from M~31 and M~33. } \label{fig:wavecor1}
\end{center}
\end{figure}

\begin{figure}
\begin{center}
\resizebox{7cm}{!}
{\includegraphics*{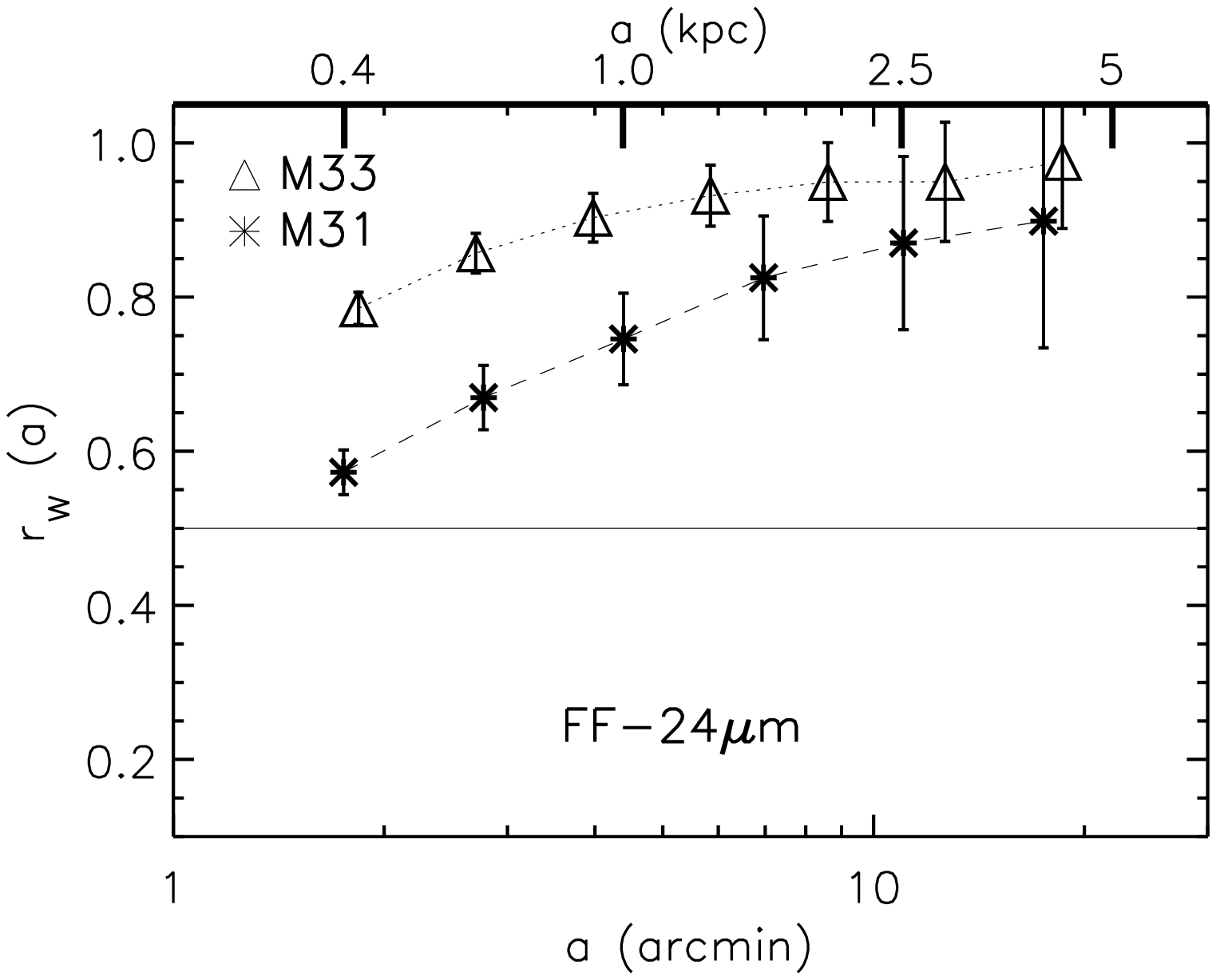}}
\resizebox{7cm}{!}
{\includegraphics*{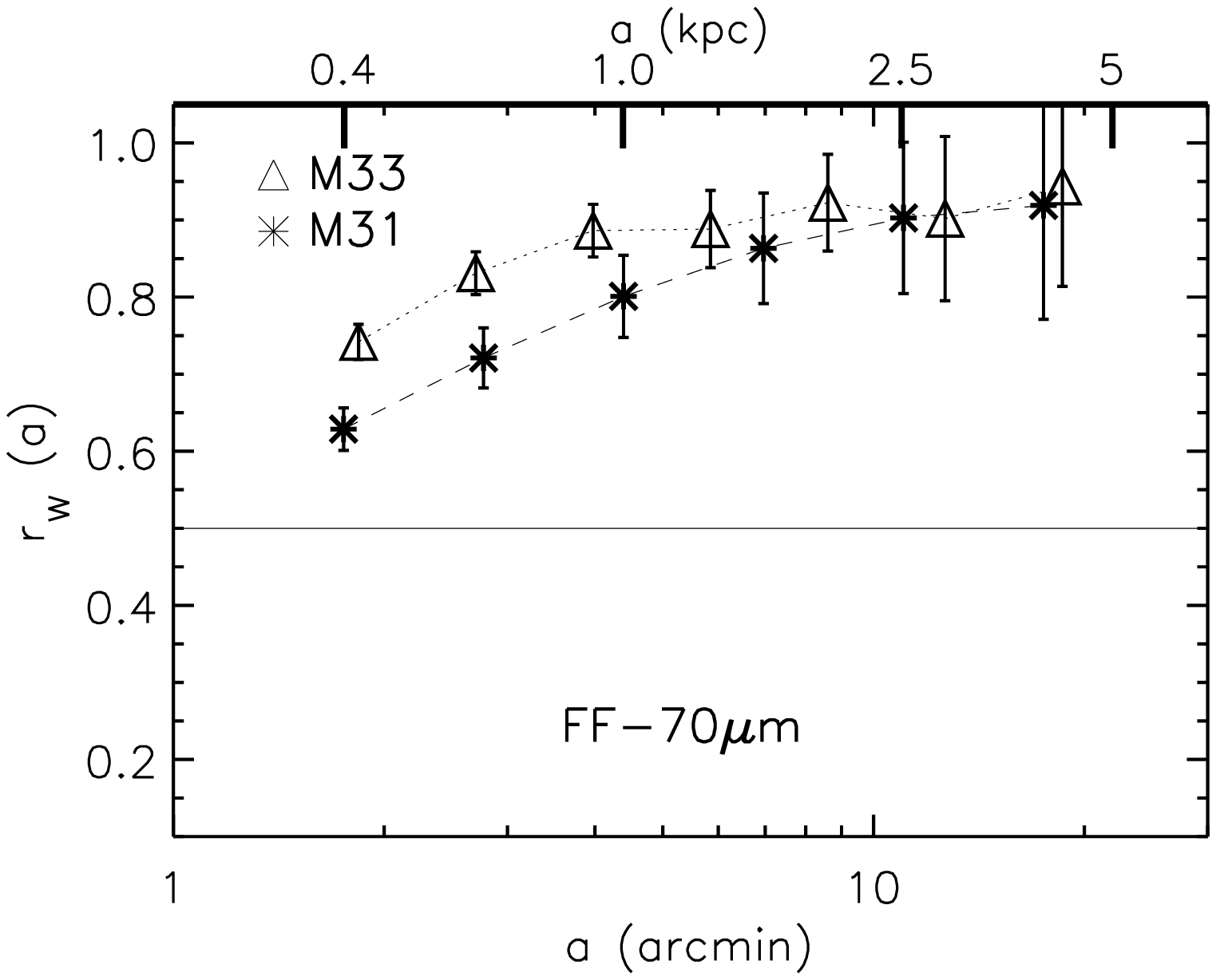}}
\resizebox{7cm}{!}
{\includegraphics*{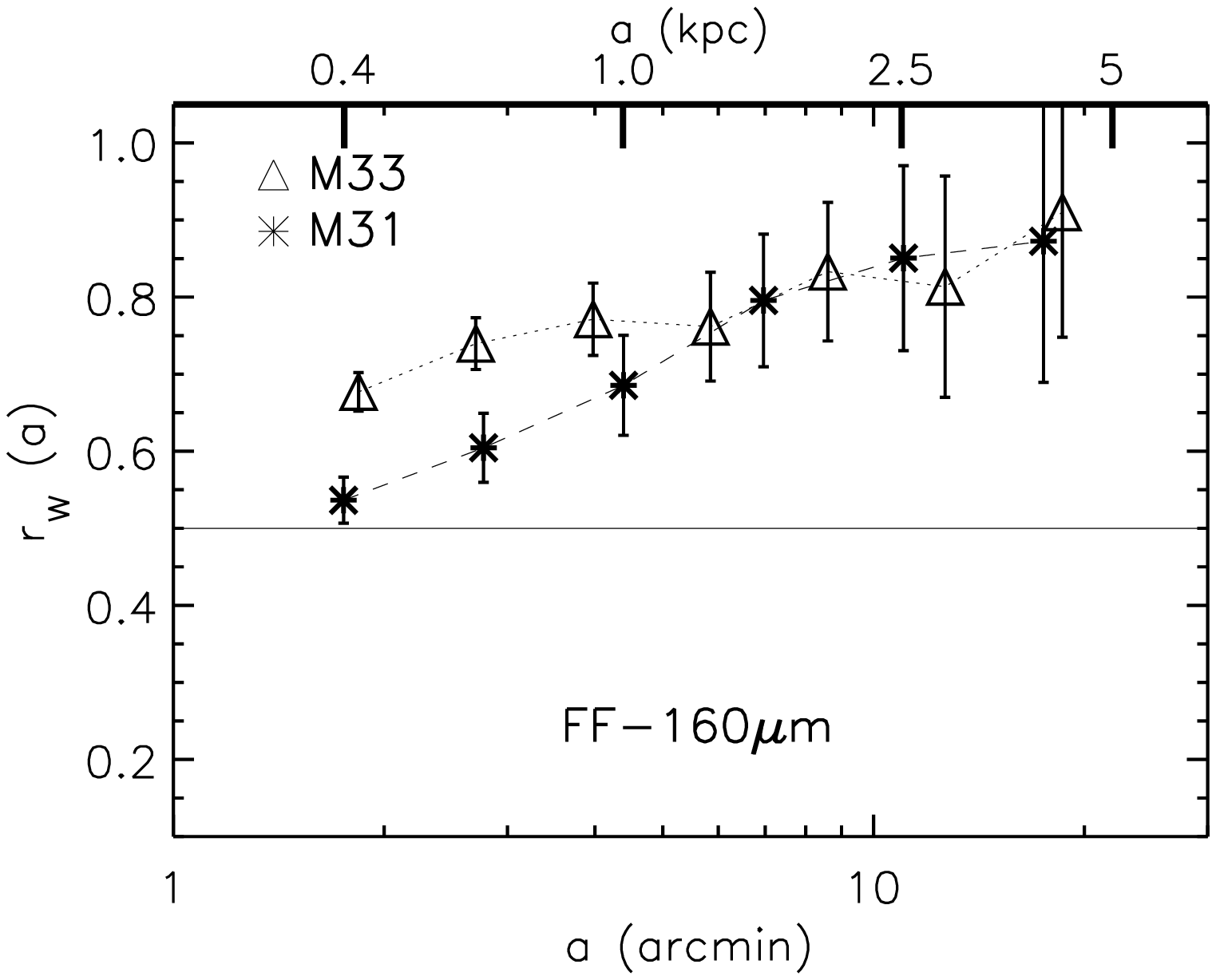}}
\caption[]{Scale-by-scale cross-correlations between IR and
20\,cm free-free emission from M~31 and M~33.} \label{fig:wavecor2}
\end{center}
\end{figure}

\begin{figure}
\begin{center}
\resizebox{7cm}{!}
{\includegraphics*{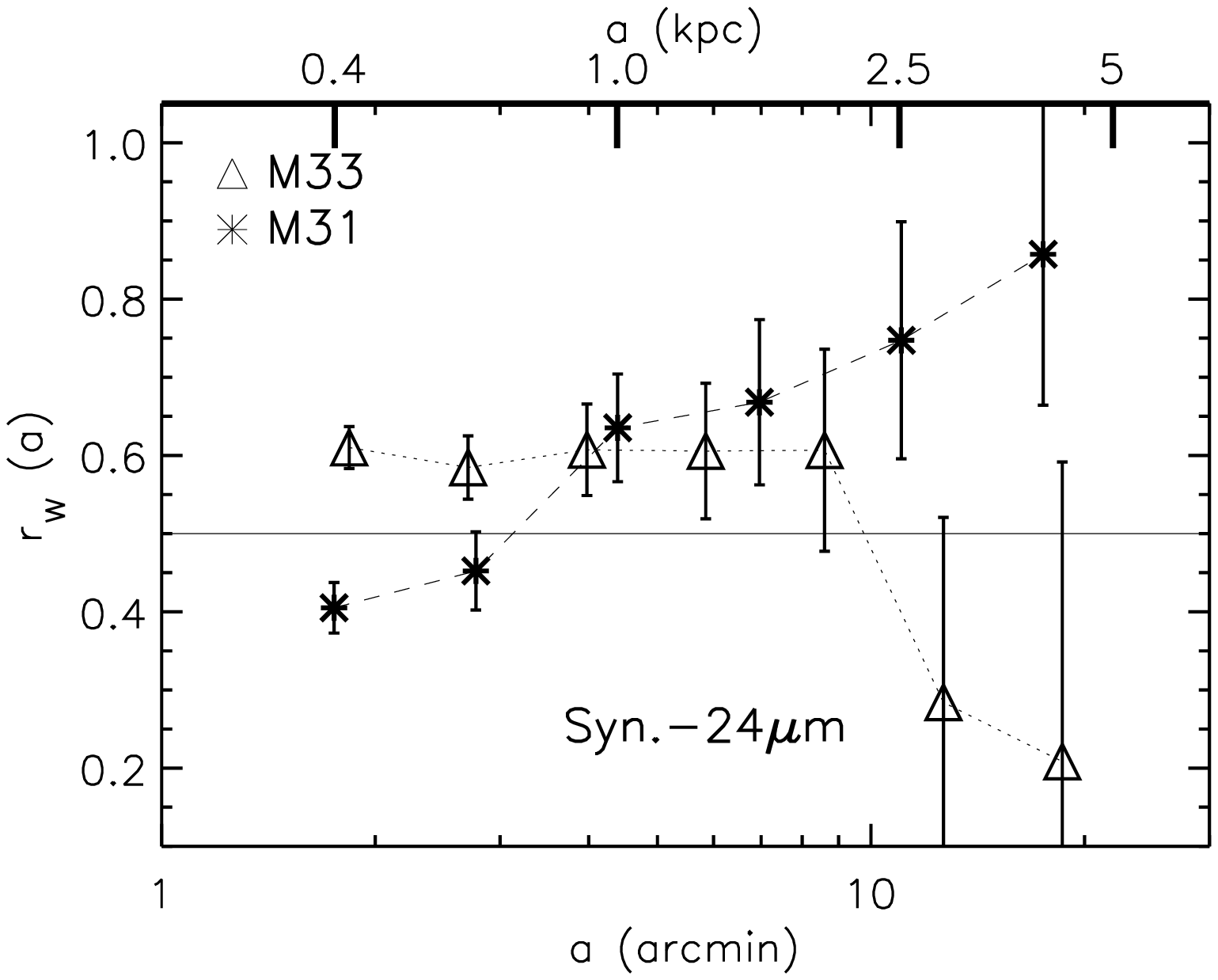}}
\resizebox{7cm}{!}
{\includegraphics*{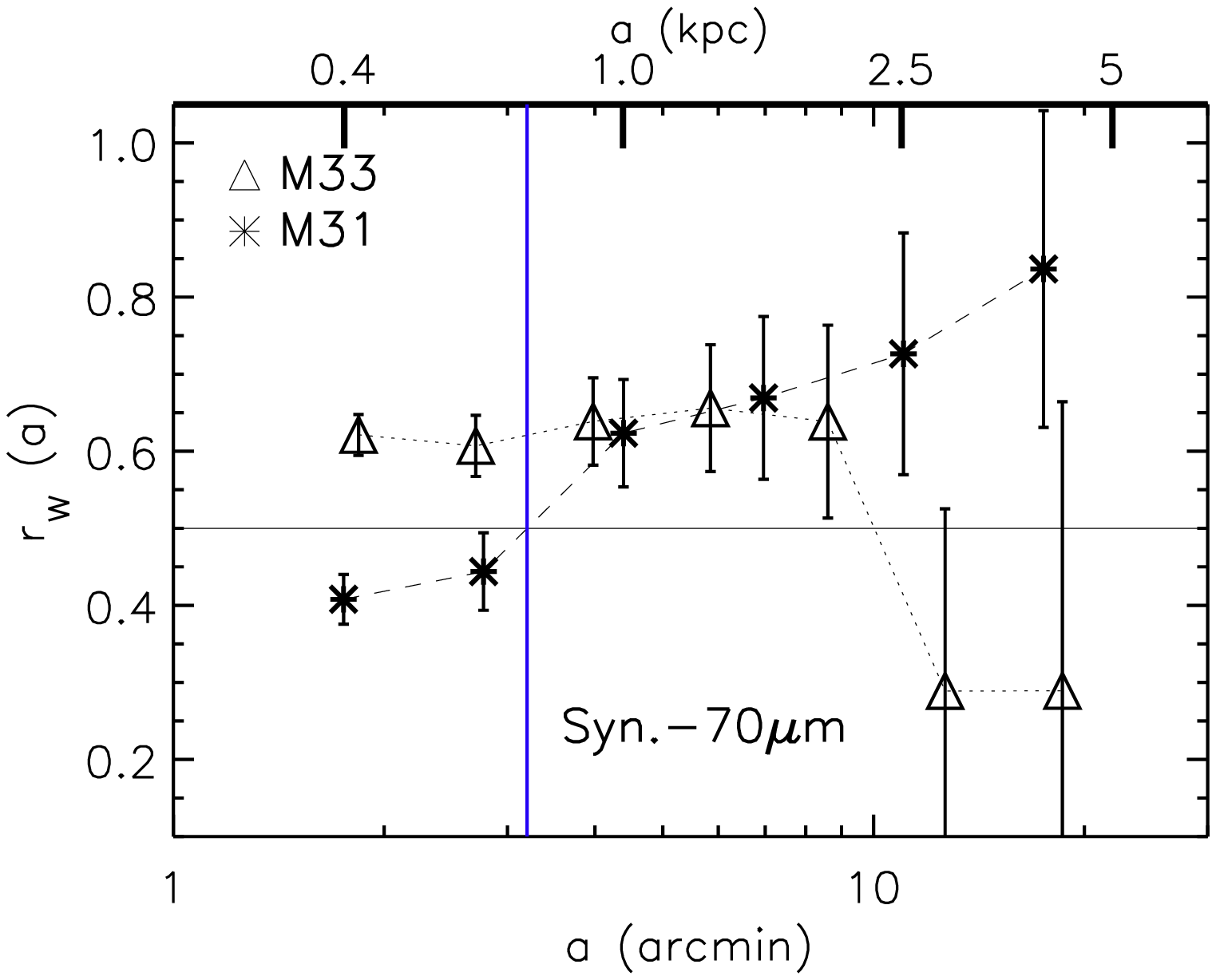}}
\resizebox{7cm}{!}
{\includegraphics*{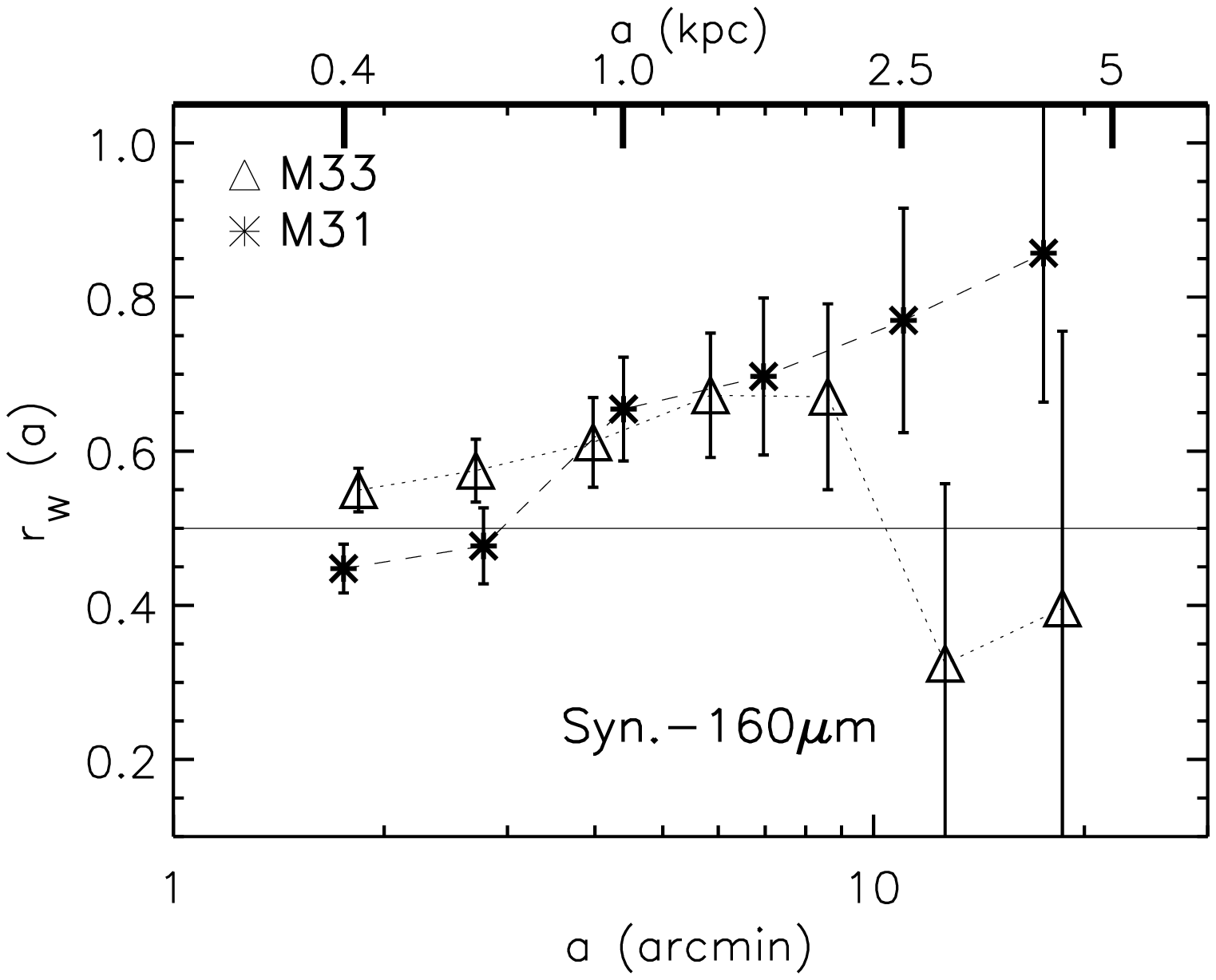}}
\caption[]{Scale-by-scale cross-correlations between IR and 20\,cm
synchrotron emission from M~31 and M~33. The vertical line in
the middle panel indicates $a=l_{\rm dif}$ for M~31.}
\label{fig:wavecor3}
\end{center}
\end{figure}

Figures~\ref{fig:wavecor1}--\ref{fig:wavecor3} show the (total)
radio--IR correlation, free-free--IR correlation and synchrotron--IR
correlation versus spatial scale in M~31 and M~33. The radio--IR
correlations  (Fig.~\ref{fig:wavecor1}) (particularly
RC--24/70\,$\mu$m) on $a<2.5$\,kpc in M~31 are weaker than in M~33,
without significant correlation ($r_w<0.5$) on the smallest scale.

The free-free and IR emissions are well correlated ($r_w > 0.5$) on
all scales in both galaxies (Fig.~\ref{fig:wavecor2}). On scales
$a \le 1.5$\,kpc, the correlations are stronger in M~33 than in M~31,
depending on the IR wavelength. 
This indicates that the importance of young O/B stars in
heating the dust is higher in M~33 than in M~31. In this early-type galaxy, the dust is largely
heated by ISRF \citep[e.g.][]{Hoernes_etal_98}, leading to a lower mean dust temperature
\citep{Tabatabaei_10} and a weaker free-free--IR correlation than in M~33. {  The free-free
emission is more tightly correlated with the 24\,$\mu$m than the 70\,$\mu$m emission in M~33,
as expected, but it is the other way around in M~31. This, along with the observed excess in the
24\,$\mu$m-to-70\,$\mu$m intensity ratio, indicates that sources other than interstellar dust heated by massive stars contribute to the 24\,$\mu$m emission in M~31 \citep[see][]{Tabatabaei_10}.}


The synchrotron--IR correlations in M~31 and M~33
(Fig.~\ref{fig:wavecor3}) are quite different. In M~31, the
correlations are good ($r_w > 0.7$) on $a>$\,2\,kpc, but decrease
towards smaller scales. On $a< 1$\,kpc, no significant correlation
($r_w < 0.5$) exists. In M~33, however, the synchrotron--IR
correlation reasonably holds ($0.5 < r_w < 0.7$) on scales  $a<
2$\,kpc, while it is weak on scales of 3--4\,kpc due to the
different distributions of synchrotron and IR emission  on these
scales (see Fig.~\ref{fig:decompose2}, bottom).
\section{Discussion}
\label{dis}
\subsection{CRE propagation length and magnetic field structure in M\,31 and M\,33}
\label{pro}
On small scales ($a<1$\,kpc), the synchrotron--IR correlation is
stronger in M~33 than in M~31, but on large scales ($a>2.5$\,kpc) it
is stronger in M~31 (Fig.~\ref{fig:wavecor3}). To understand these
differences, we look at the properties of the magnetic field
and CREs in M~31 and M~33, because the synchrotron emission is
proportional to both the strength of the total magnetic field
($B_{\rm tot}$) and the number of CREs. $B_{\rm tot}$ consists of
turbulent ($B_{\rm tur}$) and ordered ($B_{\rm ord}$) magnetic
fields ($B_{\rm tot}^2=B_{\rm tur}^2+B_{\rm ord}^2$). Since $B_{\rm
tur}$ and the number of young CREs are enhanced in SF regions, i.e.
{\it on small scales}, the synchrotron emission is also increased
\citep{chyzy,krause}, while the large-scale field $B_{\rm ord}$ is
not enhanced in SF regions \citep{Frick_etal_01,chyzy}. Hence
one may speculate that a higher SF rate per unit area
($\Sigma_{SFR}$)  like  in M~33 (see Table~3) improves the
synchrotron--IR correlation on small scales. However,
\cite{Boulanger_88} did not find a correlation between radio and IR
emission for {  the star-forming Orion nebula of a few 100\,pc}.
As the Milky Way also has a higher $\Sigma_{SFR}$ than M~31, this is
inconsistent with the above speculation.
%

The differences between the correlations in M~31 and M~33 are
better explained by differences in the propagation of CREs. On our
smallest scale of $a=0.4$\,kpc, the reasonable synchrotron--IR
correlation in M~33 implies that the structures of the two {  intensity maps}
are similar and that the positions of maxima and minima on the sky
largely agree. This is visible in Fig.~\ref{fig:decompose} (bottom),
where more or less the same bright sources (which are complexes of
SF regions) can be traced in the synchrotron and the 160\,$\mu$m
maps. This means that CREs are still near their places of origin in
the SF regions. Thus, the propagation length of CREs ($l_{\rm
prop}$) in M~33 is smaller than 0.4\,kpc. In M~31, however, the
insignificant correlation ($r_w < 0.5$) on $a < 1$\,kpc implies
that the structures of synchrotron and dust emission differ.
Fig.~\ref{fig:decompose} (top) shows this difference: the dust
emission mainly emerges from the dust lanes and SF regions in the
spiral arms, whereas the synchrotron emission comes from the bright
diffuse ring. This indicates that the bulk of the CREs have
propagated away from their places of origin in the SF regions.
Since the synchrotron--IR correlation holds ($r_w \ge 0.5$) on
scales of $a\gtrsim 1$\,kpc, we conclude that the CREs must have
propagated over a length of $l_{\rm prop}\simeq 1$\,kpc \footnote{The kpc
scales used here refer to the major axis.}.
Hence, $l_{\rm prop}$ in M~31 is larger than in M~33. Possible
reasons for this difference are discussed below.

The synchrotron--IR correlation on large scales ($a>2.5$\,kpc) is
due to diffuse emission associated with the `10\,kpc ring' in M~31
and with the central region in M~33. Because $l_{\rm prop}$ in M~31
is larger than in M~33, the synchrotron emission will be more
diffuse, leading to a better correlation with the diffuse dust
emission in M~31. The bright  HII region complex NGC\,604 and
the central disk seen in the dust emission from M~33 on scales
3--4\,kpc (e.g. see Fig.~\ref{fig:decompose2}) have no counterpart
in the synchrotron emission, which causes the low correlation
coefficients on $a=3-4$\,kpc. This could be due to the combined
effect of the small $l_{\rm prop}$ in the disk that keeps the CREs
near SF regions, and the fact that the large-scale $B_{\rm tot}$ is
not confined to the thin disk in M~33.
\citet{Tabatabaei_08} found that M~33 has a vertical magnetic field
extending to $R\approx5$\,kpc, which is strongest in the
central disk. This causes propagation of CREs from  the SF
regions in the thin disk into the halo, as observed in edge-on
galaxies \citep [e.g.][]{Heesen_09}. CRE propagation along the
vertical field into the halo reduces the synchrotron intensity in
the SF regions {  \citep[see also][]{Tabatabaei_2_07}}. This further
weakens the synchrotron--IR correlation because the dust emission is
concentrated in the thin disk.

In M~31, since $B_{\rm tot}$ is
confined to the dusty thin disk \citep[no synchrotron
halo,][]{Moss,Fletcher_04}, the better correlations on large scales are due to the
similar structures in the thin disks (as shown in
{  Fig.~6).  This suggests coupling of the magnetic field to the gas
mixed with the dust, which could explain the synchrotron--IR correlation
in M~31, as proposed by \cite{Hoernes_etal_98}.}

\begin{figure}
\begin{center}
\resizebox{7cm}{!}{\includegraphics*{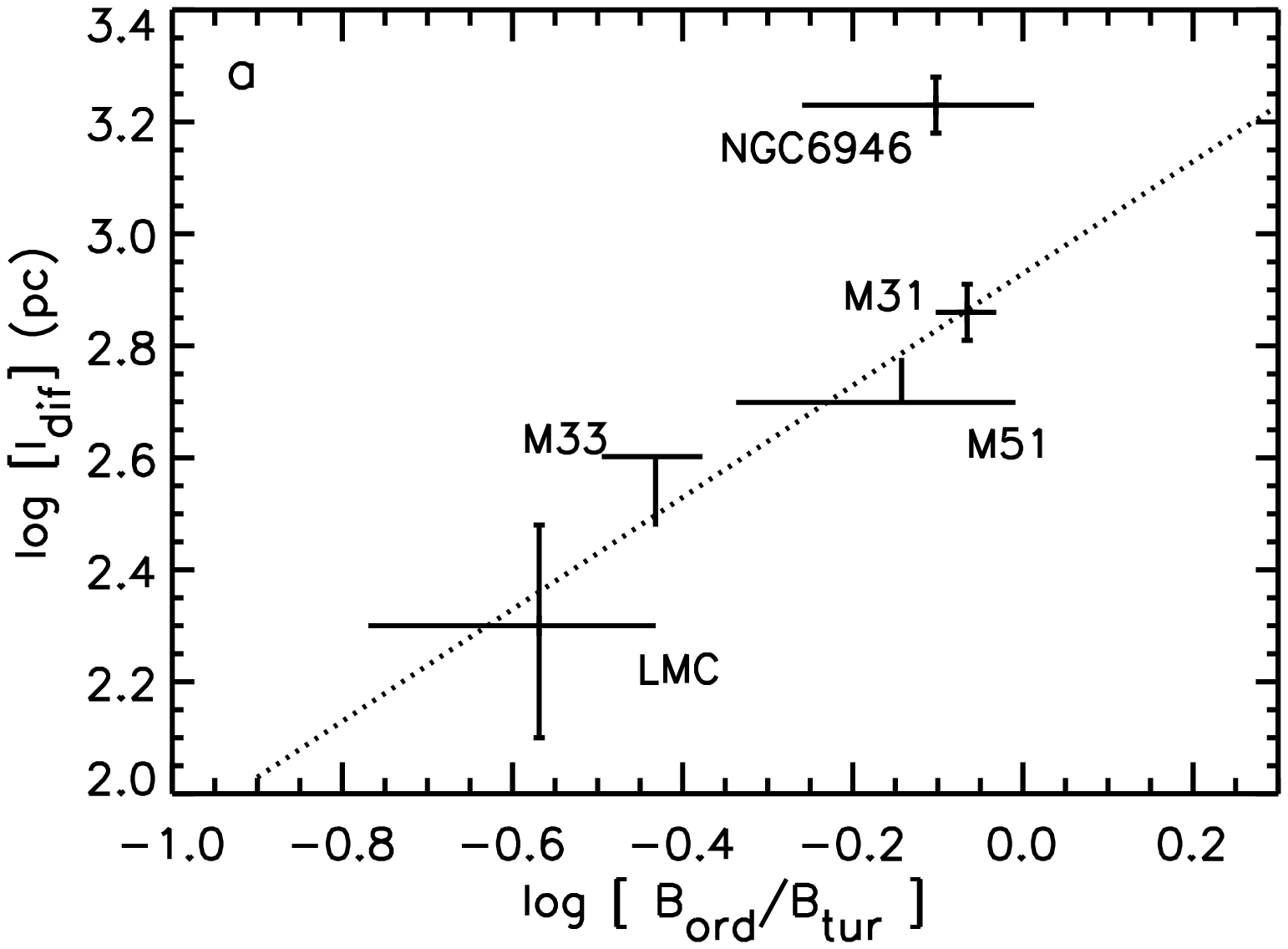}}
\resizebox{7cm}{!}{\includegraphics*{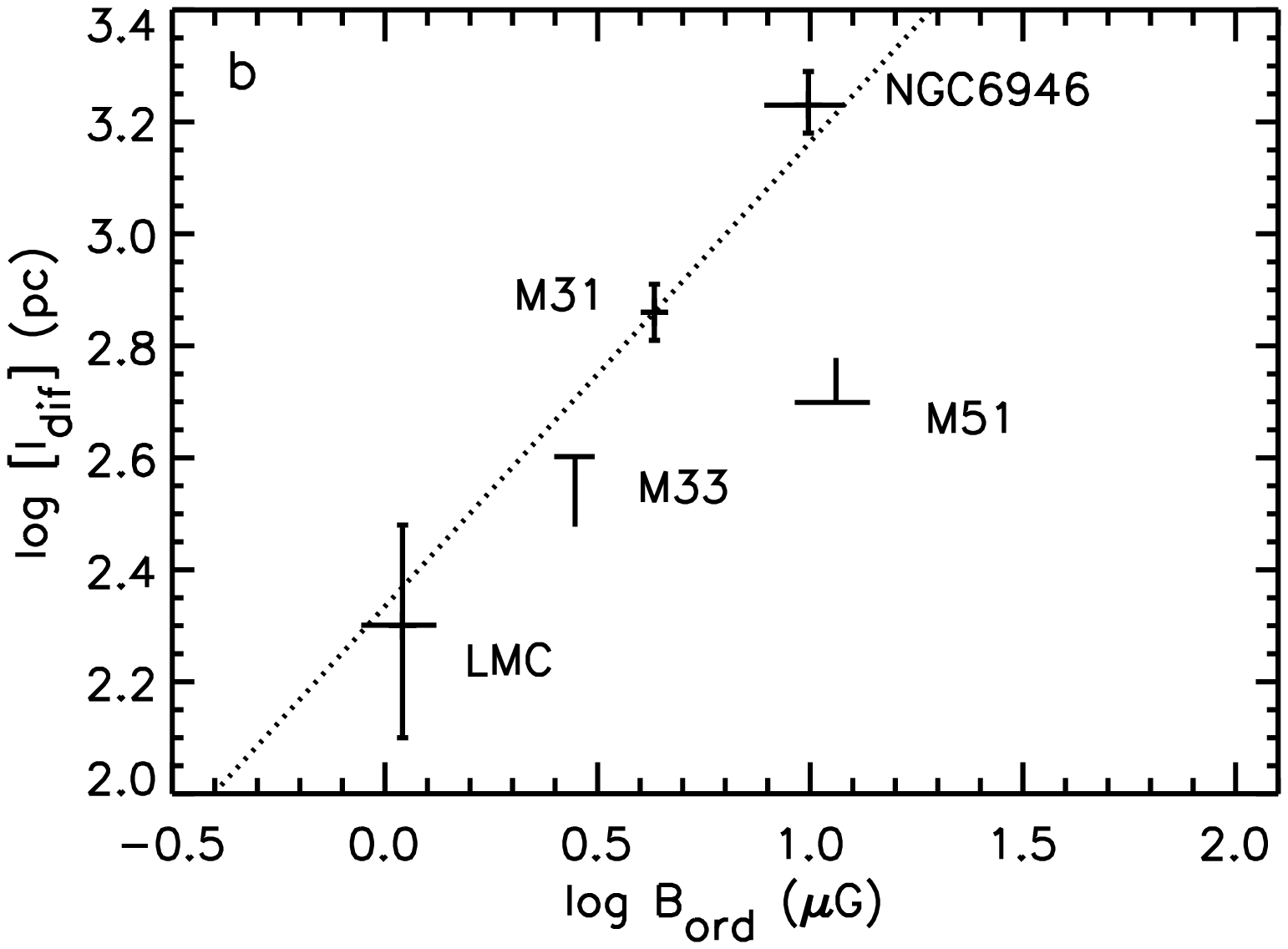}}
\resizebox{7cm}{!}{\includegraphics*{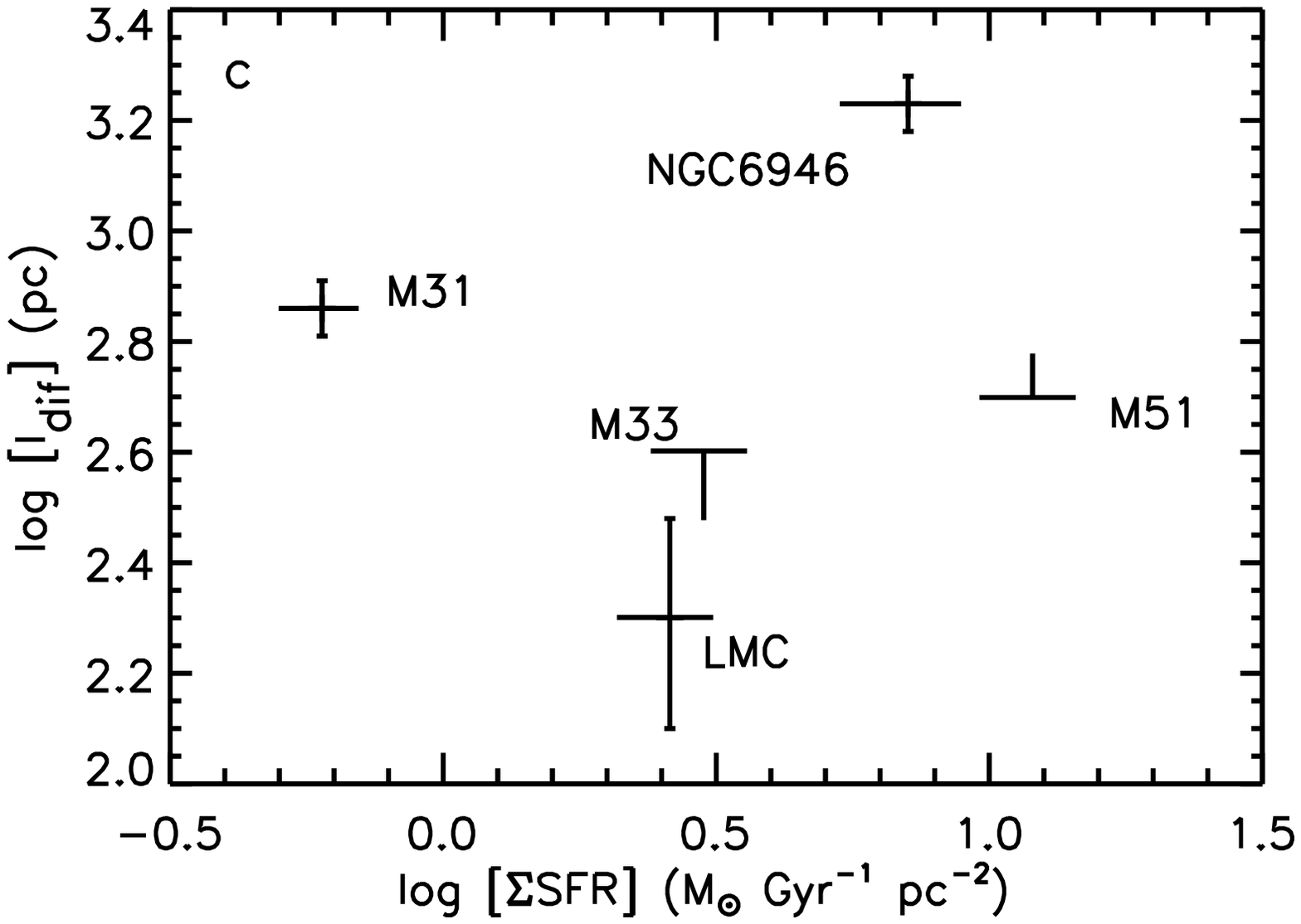}}
\resizebox{7cm}{!}{\includegraphics*{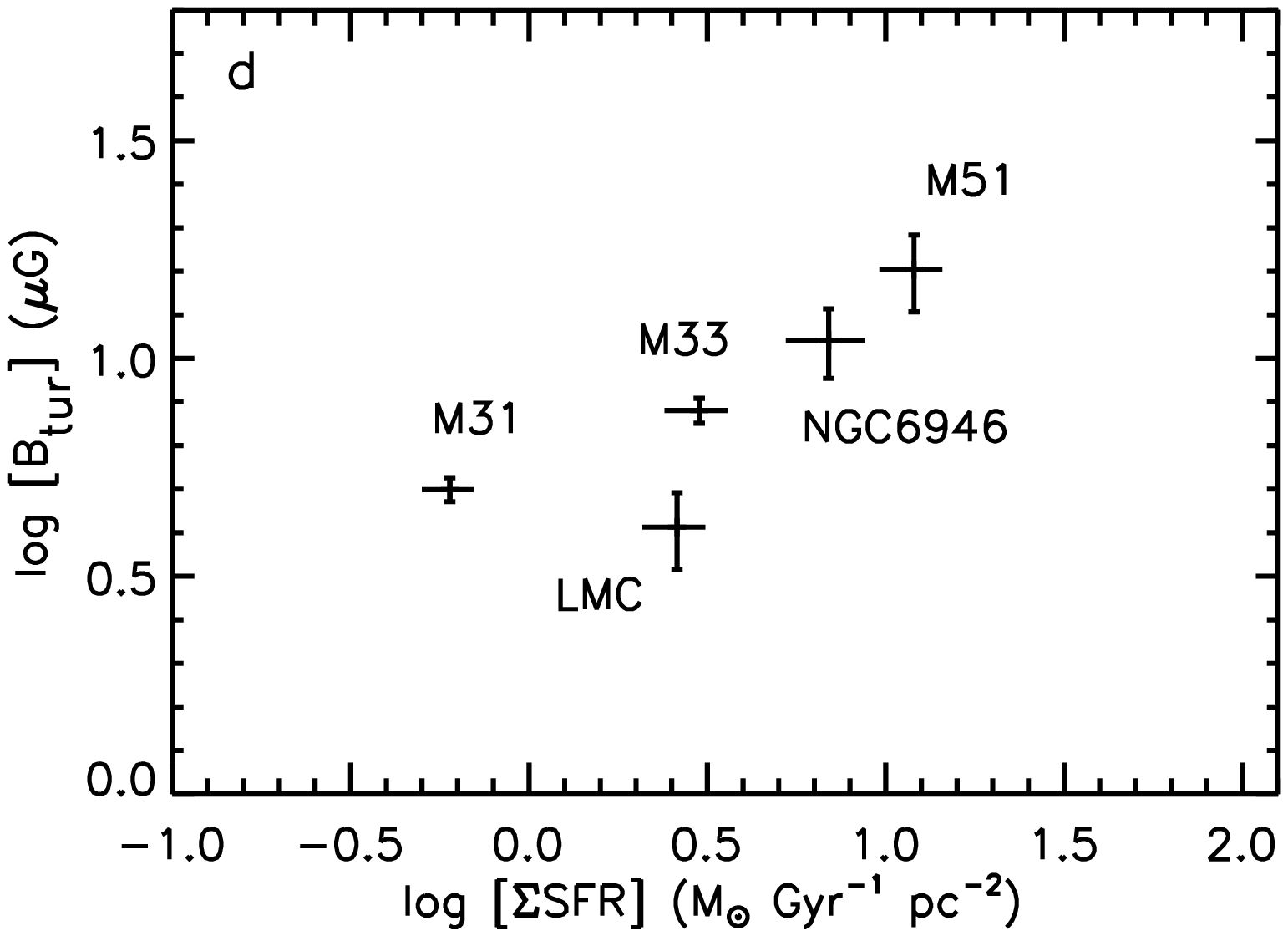}}
\caption[]{  The smallest scale on which the RC and IR emissions are
correlated. From top to bottom $l_{\rm dif}$ versus B$_{\rm ord}$/B$_{\rm tur}$ ({\it a}), B$_{\rm ord}$ ({\it b}),
and $\Sigma_{\rm SFR}$  ({\it c}). Also shown is B$_{\rm tur}$ versus $\Sigma_{\rm SFR}$ ({\it d}). For the LMC and M~51, 20\%
uncertainty is assumed. The dotted lines in ({\it a}) and ({\it b}) are not fits, but show the relation expected from Eq.~13
with slope=1 in ({\it a})  and slope=5/6 in ({\it b}), adjusted to M~31.}
\label{fig:co}
\end{center}
\end{figure}

\subsection{CRE propagation models}
\label{role}
As indicated above, the difference in $l_{\rm prop}$ between
M~31 and M~33 could be related to their different magnetic field
properties. The degree of order of the magnetic field plays an important
role in the propagation of CREs, both in models of  streaming
instability \citep[e.g. ][]{Kulsrud,Ensslin} and in diffusion
models \citep[e.g. ][]{Chuvilgin,Breitschwerdt,Dogiel,Shalchi}.
A larger ordering, i.e. a larger $B_{\rm ord}/B_{\rm tur}$,
helps CREs to propagate over larger distances  along the ordered
magnetic field resulting in a larger $l_{\rm prop}$. This is in
agreement with the observational results: $B_{\rm ord}/B_{\rm tur}$
is indeed larger in M~31 than in M~33 (see Table~3), as the larger
$l_{\rm prop}$ suggests. This offers a new way to estimate $l_{\rm
prop}$ from the synchrotron--IR correlation in galaxies.

{  In the case of streaming, CREs move along the lines of the ordered
field with the Alfv{\'e}n velocity ($v_{\rm A}$) and propagate over
a distance $l_{\rm prop}= v_{\rm A} \, t_{\rm CRE}$ during their
lifetime ($t_{\rm CRE}$). Observations indicate equipartition
between the kinetic and magnetic energy densities, so that
$v_{\rm A} \approx constant$.
If the CRE lifetime is limited by synchrotron loss,
$t_{\rm CRE} = t_{\rm syn} \propto B_{\rm tot}^{\,\,-1.5} \, \nu^{\,\,-0.5}$,
where $\nu$ is the frequency at which the CREs of energy E radiate
\footnote {$\nu \propto B_{\rm tot} \, E^2$}
\citep[neglecting inverse Compton losses, which are smaller than synchrotron losses;][]{Murphy_09}, we get $l_{\rm prop} \propto B_{\rm tot}^{\,\,-1.5} \,
\nu^{\,\,-0.5}$. If $t_{\rm CRE}$ is given by the time that the CREs reside in the
thin galactic disk until they move into the thick disk/halo, the confinement time ($t_{\rm conf}$),
then $l_{\rm prop} \propto t_{\rm conf}$.

However, as streaming needs longer stretches of the ordered field
to be effective, it may not be important in the turbulent thin disk
of galaxies \citep{Dogiel,Ensslin}. Furthermore, there is evidence that in the Milky Way
on scales $ < 1$\,kpc CRE propagation is isotropic \citep{Strong_07,Stepanov_12},
which is inconsistent with streaming. In the following, we assume that generally
CRE diffusion dominates, except in regions with highly ordered fields, where streaming
may dominate.}

In the case of diffusion, the particles are scattered by the
irregularities in the turbulent magnetic field. The diffusion length ($l_{\rm
dif}$) depends on the diffusion coefficient ($D$) and {  $t_{\rm CRE}$
as $l_{\rm dif} = (D \, t_{\rm CRE})^{0.5}$,} where $D \propto D_{\rm
E} \, E^{0.5} \propto D_{\rm E} \, \nu^{0.25} \, B_{\rm
tot}^{\,\,-0.25}$, {  with $D_{\rm E}$ being the energy dependent
diffusion coefficient.} Models of diffusive CR propagation in the ISM
indicate that $D_{\rm E} \propto (B_{\rm ord}/B_{\rm tur})^2$
\citep [][]{Chuvilgin,Breitschwerdt,Yan,Dogiel,Shalchi}.
{  Assuming that diffusion along $B_{\rm ord}$
dominates, the parallel diffusion coefficient is $D_{\parallel} \propto (B_{\rm ord}/B_{\rm tur})^2 \, B_{\rm ord}^{\,\,-1/3}$
\citep[][Eq.~3.41]{Shalchi}. Hence, observing at a fixed frequency,
we expect that during their synchrotron lifetime $t_{\rm syn}$ the CRE will diffuse over the pathlength:
\begin{equation}
l_{\rm dif} \propto (B_{\rm ord}/B_{\rm tur}) \, B_{\rm ord}^{\,\,-1/6} \, B_{\rm tot}^{\,\,-7/8},
\label{dif1}
\end{equation}
where we used $t_{\rm syn} \propto B_{\rm tot}^{\,\,-1.5}$.

If, on the other hand, the period that the CREs spend in the galactic disk is the confinement time, then:
\begin{equation}
l_{\rm dif} \propto (B_{\rm ord}/B_{\rm tur}) \, B_{\rm ord}^{\,\,-1/6} \, B_{\rm tot}^{\,\,-1/8}\, t_{\rm conf}^{1/2}.
\label{dif2}
\end{equation}
}

%
%
%

For M~31 we estimate $l_{\rm dif}=730\pm90$\,pc, where $l_{\rm dif}$
is the smallest scale on which the synchrotron--IR correlation is
found to hold ($r_w=0.5$) \footnote{This is determined using the
synchrotron--70\,$\mu$m correlation, as in M~31 the 70\,$\mu$m
emission traces SF and CREs sources better than other IR bands
(Sect.~\ref{algorithms}).}. The smallest scales contain the SF
regions, where most CREs are young \citep[as suggested
by][]{Murphy_06}, and $B_{\rm tot}$ is larger than average $B_{\rm
tot}$. As the synchrotron emission from the smallest scale in M~33
is about 1.9 times that in M~31 (Sect.\,4.1), {  we estimate that
$B_{\rm tot}$ is about $(1.9)^{1/(1+\alpha_{n})}\simeq 1.5 $ times stronger
\citep[assuming equipartition between the magnetic field and CREs energy
densities and using a nonthermal spectral index of $\alpha_{n}\simeq 0.6$,][]{Tabatabaei_2_07}.
With the smaller value of $B_{\rm ord}/B_{\rm tur}$ and  $B_{\rm ord}$ (see
Table~3), and equal $t_{\rm conf}$ (see Sect. \ref{m51}), this leads  to $l_{\rm dif} = 320\pm70$\,pc
in M~33, in agreement with our estimate of $< 400$\,pc from the synchrotron--IR
correlation. For comparison, using the somewhat lower mean
value of $B_{\rm tot}$ given in Table~3 yields  $l_{\rm dif} = 330\pm70$\,pc.}
%
%

So far we have measured $l_{\rm dif}$ in the sky plane, i.e.
along the major axis of the galaxy and perpendicular to it. We may
estimate $l_{\rm dif}$ in the disk plane by assuming that the
diffusion lengths along the major axis ($l_{\rm x}$) and the minor
axis ($l_{\rm y}$) are the same, as is indicated by the observed
isotropy of CREs {  in the Milky Way} on scales $<1$\,kpc  \citep{Strong_07,
Stepanov_12}. The mean diffusion length along the plane then is
$l_{\rm dif,xy} = \sqrt{l_{\rm x}^2+l_{\rm y}^2} = l_{\rm dif} \,
\sqrt{2}$.

\subsection{Comparison with other galaxies}
\label{m51}
{  NGC\,6946} and the LMC  are the only two nearby galaxies, for which
similar scale-by-scale studies of the {  synchrotron--IR correlation
exist \citep[][Hughes et al. in prep.]{Taba_13}, and it is
interesting to compare the five galaxies.
 Table~3 shows all values of $l_{\rm dif}$, the smallest scale on which the synchrotron and IR emission
are correlated. For M~51, \cite{Dumas} presented the radio continuum-IR correlation as a function of scale.
The smallest scale on which the radio continuum (including both synchrotron and free-free emission) and
IR emission are correlated is a lower limit for $l_{\rm dif}$ (also given in Table~3), because the good
free-free--IR correlation increases the radio--IR correlation on small scales. }



{  We first tested if $l_{\rm dif}$ indeed follows Eq.~(\ref{dif1}) given above,
but did not find such a correlation, especially not an inverse dependence on $ B_{\rm tot}$.
This indicates that the synchrotron lifetime
$t_{\rm syn}$ does not determine the CRE lifetime because the confinement
time $t_{\rm conf}$ is shorter than $t_{\rm syn}$, which was also found by \citet{Lacki_10}.
Interestingly, we do find a clear dependence on $(B_{\rm ord}/ B_{\rm tur})$
(see Fig.~\ref{fig:co}a), which is consistent with Eq.~(\ref{dif2})
\footnote{  The correlation with the full Eq. (13) is somewhat worse than that with  $(B_{\rm ord}/ B_{\rm tur})$
alone, but still consistent with the data. The extra factors $B_{\rm ord}^{-1/6}$ and $ B_{\rm tot}^{-1/8}$ apparently
worsen the correlation with $B_{\rm ord}/B_{\rm tur}$.}.
The fact that this dependency exists shows that the confinement time in the five
galaxies will be approximately the same. This is understandable because the size of star
formation regions and the scale height of the thin disk \citep[typically 0.3\,kpc,][]{krause}
are about the same in galaxies, so the CREs need about the same time to diffuse out of the disk.

NGC\,6946 has ``magnetic arms'' where the magnetic field is very regular \citep{Beck_07,Taba_13}.
The fact that the point for NGC\,6946 in Fig.~\ref{fig:co}a is above the dotted
line may indicate that along these arms CRE streaming occurs, while diffusion dominates elsewhere.
It is also possible that $t_{\rm conf}$  in NGC\,6946 is very long because CREs in the magnetic arms
will be more strongly confined to the thin disk than in the spiral arms. A longer $t_{\rm conf}$
is consistent with the conclusion of \citet{Taba_13} that $l_{\rm dif}$ in NGC~6946 agrees with
the diffusion length expected from synchrotron and Inverse Compton losses during $t_{\rm syn}$ ($t_{\rm conf} \sim t_{\rm syn}$).

Since between the galaxies the values of $B_{\rm tur}$ vary less than those of $B_{\rm ord}$,
Eq.~(\ref{dif2}) may be approximated by $l_{\rm dif} \propto B_{\rm ord}^{\,\,5/6}$. Fig.~\ref{fig:co}b shows that this is also consistent with our data.}

%

We note that $l_{\rm dif}$ may also depend on the
star-formation rate (SFR) per unit area ($\Sigma_{\rm SFR}$). In
galaxies with high $\Sigma_{\rm SFR}$ the filling factor of young
CREs on small scales (i.e. in SF regions) will be higher than in
galaxies with low $\Sigma_{\rm SFR}$, which could lead to a smaller
{  $l_{\rm dif}$ than that of the bulk of CREs.} \cite{Breitschwerdt}
showed that a higher $\Sigma_{\rm SFR}$ favors convective
propagation of CREs from the SF regions into the halo, which reduces
the time that CREs spend in the thin disk and will decrease
{  $l_{\rm dif}$. Furthermore, $B_{\rm tur}$ increases in SF regions and
could lead to a lower $l_{\rm dif}$. Fig.~\ref{fig:co}d shows
that $B_{\rm tur}$ is indeed correlated with $\Sigma_{\rm SFR}$
\citep[see also][]{chyzy,krause,Taba_13}, but $l_{\rm dif}$ is
not (Fig.~\ref{fig:co}c).
%
%
Instead, $l_{\rm dif}$ scales with the ratio $B_{\rm ord}$/$B_{\rm tur}$. 
The reason for {  the lack of} correlation between $l_{\rm dif}$ and $\Sigma_{\rm SFR}$  
could be the weak dependence of $B_{\rm tur}$ on $\Sigma_{\rm SFR}$
\citep[with an exponent of about 0.2 within galaxies,][ and about 0.6 between galaxies, see Fig.~\ref{fig:co}d]{chyzy,Taba_13}. 
Moreover, $B_{\rm ord}$ is not related to the SFR  \citep{Frick_etal_01,chyzy,krause,Taba_13},  but is
controlled by various dynamical and environmental effects \citep{Fletcher_11}.
Thus it is expected that $l_{\rm dif}$ does not directly scale with $\Sigma_{\rm SFR}$.
Hence, although the presence of massive star formation influences
the field structure, it is the ratio $B_{\rm ord}$/$B_{\rm tur}$ that is crucial for the
propagation of the CREs.}



%
\subsection{Comparison with smoothing experiments}
\label{m08}

{  \citet{Murphy_08} determined propagation lengths of CREs in 18 well-resolved
galaxies ($< 1\,$kpc resolution) by smoothing 70\,$\mu$m Spitzer images until they
matched the 22\,cm Westerbork maps, corrected for thermal emission. Using a wavelet
technique, they separated the 70\,$\mu$m images in a ``structure component'' of scales $< 1$\,kpc
containing the star formation regions and a diffuse ``disk component'' of scales $> 1$\,kpc,
heated by older stars. By separately smoothing these components with exponential kernels,
until their sum matched the corrected radio image, they determined a structure scale length
and a disk scale length, which they interpreted as diffusion lengths of two populations
of CREs. There are some major differences between the approach of \citet{Murphy_08} and
our work, making a comparison of results difficult.

1. \citet{Murphy_08} assume that supernova remnants (SNRs) located in the structure component as well as re-acceleration in ISM shock waves in the disk component are sources of CREs. However, re-acceleration of CREs in the ISM has not been proven to exist -- neither theoretically nor observationally.
In our work, we assume that all CREs originate from SNRs.

2. \citet{Murphy_08} claim that the CREs in the structure component have
been accelerated only recently and have a typical age of $\gtrsim 4$
times shorter than the CREs in the disk component. In our work, we
assume that that CRE acceleration time is much shorter than the CRE
radiation time and we argue that the radiation time is limited by the
confinement time in the thin disk.

3. \citet{Murphy_08} ignore the role of the magnetic fields in diffusion 
of CREs.
Their calculations are based on some assumed values of the total magnetic field strength. In our work, we show the importance of the magnetic fields with respect to strength and ordering. We use the magnetic field strengths independently measured for each galaxy.

4. Both the structure and disk scale lengths derived by \citet{Murphy_08} decrease with increasing $\Sigma_{\rm SFR}$, especially for galaxies with high $\Sigma_{\rm SFR}$. They attribute it to the increasing importance of the structure component that contains the SF regions.
However, our data do not show a dependence of $l_{\rm dif}$ on $\Sigma_{\rm SFR}$
(see Fig.~\ref{fig:co}, lower middle panel), although our range of $\Sigma_{\rm SFR}$ is larger than that of the sample of \citet{Murphy_08}.

{  Our} main concern about the work of \citet{Murphy_08} is the reality of the disk
component as a CRE source distribution. We believe that their scale $l_{\rm disk}$ is the difference between the exponential radial scale lengths of the diffuse dust emission powered by old stars and the extended synchrotron emission (see their example for NGC~6946). Since old stars are not progenitors of SNRs, they cannot produce CREs. Moreover, if there really would be two types of CRE sources, this should cause two peaks in the wavelet cross-correlation function, one at small and one at large scales. Something like that may be visible in NGC~6946 \citep{Taba_13}, but clearly not in M~31 and M~33. It seems unlikely that the ``diffuse source component'' exists.

\citet{Murphy_08} also derived global scale lengths, $l_{\rm glob}$, after smoothing of
the unseparated, observed 70\,$\mu$m image,
 and it is interesting to compare their values
for M~51 and NGC~6946 with our $l_{\rm dif}$. For M~51, they find $l_{\rm glob} =
500\pm\,100$\,pc which is consistent with our lower limit of $l_{\rm dif} > 500$\,pc.
For NGC~6946, however, they derive $l_{\rm glob} = 600\pm\,140$\,pc, much lower than our
$l_{\rm dif} = 1700\pm\,200$\,pc. {  Recently, \cite{Murphy_12} applied the smoothing method to
the LMC ($R<2.7\arcmin$) and found a propagation length of 200-400\,pc,
depending on the smoothing kernel used. This agrees with the value
for the LMC (without 30 Dor) of $l_{\rm dif}=200\pm90$\,pc \citep[][and in prep.]{Hughes_etal_06} in Table~3. However, as there is no general} consistency between the results of the two methods, a thorough comparison for a sample of galaxies with  low and high $\Sigma_{\rm SFR}$, based on the same data, will be needed to better understand the differences.}

\begin{table*}
\begin{center}
\caption{  Magnetic field strengths, $\Sigma_{\rm SFR}$, and $l_{\rm
dif}$ in 5 nearby galaxies.}
\begin{tabular}{l l l l l l l}
\hline
Galaxy  & B$_{\rm tot}$ & B$_{\rm ord}$ &B$_{\rm tur}$ & B$_{\rm ord}$/B$_{\rm tur}$ & $\Sigma_{\rm SFR}$ & $l_{\rm dif}$ \\
& ($\mu$G)   & ($\mu$G)          & ($\mu$G)          & & (M$_{\odot}$Gyr$^{-1}$pc$^{-2}$) & (pc) \\
\hline
M~31          & 6.6\,$\pm$\,0.3  & 4.3\,$\pm$\,0.3 & 5.0\,$\pm$\,0.2  & 0.86\,$\pm$\,0.07 & 0.6\,$\pm$\,0.1 &
 730\,$\pm90$ \\
M~33          & 8.1\,$\pm$\,0.5  & 2.8\,$\pm$\,0.3 & 7.6\,$\pm$\,0.5  & 0.37\,$\pm$\,0.05 & 3.0\,$\pm$\,0.6 &
$<$\,400  \\
NGC\,6946     & 16.0\,$\pm$\,1.5 & 9.9\,$\pm$\,2.1 & 12.6\,$\pm$\,2.7 & 0.79\,$\pm$\,0.24 & 7.1\,$\pm$\,1.8 & 1700\,$\pm$\,200  $^{1}$    \\
LMC & 4.2              & 1.1 $^{2}$      & 4.1 $^{2}$       & 0.27\,$\pm$\,0.10 & 2.6 $^{3}$      &
200\,$\pm90$ $^{4}$ \\
M~51          & 19.7             & 11.5 $^{5}$     & 16.0 $^{5}$      & 0.72\,$\pm$\,0.26 & 12 $^{6}$       &
$>$\,500 $^{7}$ \\
\hline
\end{tabular}
\end{center}
Note that apart from $l_{\rm dif}$ all values refer to
$30\arcmin<R<50\arcmin$ in M~31, $R<21\arcmin$ in M~33,
$R<3.9\arcmin$ in M~51, and $R<3.8^{\circ}$ in the LMC.
{  The 30\,Dor region and the star-forming ridge are not considered for the LMC. For M51, $l_{\rm dif}$ is determined from the total radio continuum--IR correlation (see text).}\\
\\
1. \citet{Taba_13}\\
2. \citet{Gaensler_05}\\
3. \citet{Whitney_08}, revised for a Salpeter IMF\\
4. \citet{Hughes_etal_06} and in prep. \\
5. \citet{Fletcher_11}\\
6. \citet{Leroy}\\
7. \citet{Dumas}\\

\end{table*}

\section{Summary and conclusion}
Separating the 20\,cm free-free and synchrotron emission, we find
bright synchrotron emission close to the sites of star formation,
concentrated in a smooth wide ring in M~31 and in a smooth
spiral pattern in M~33. On the resolved scale of star forming
regions, $\simeq 200$\,pc, both the free-free emission and the
synchrotron emission are stronger in M~33 than in M~31. The wavelet
analysis shows that in M~31 and M~33 the free-free emission is
similarly distributed versus scale (i.e. dominant on small scales,
$a\lesssim 2$\,kpc), while the synchrotron emission behaves
differently: it is dominant on large scales ($\simeq$\,4\,kpc) in
M~31, unlike in M~33.

Using the Spitzer MIPS data at 24, 70, and 160\,$\mu$m, we derive
radio--IR correlations as a function of scale, separately for
free-free and synchrotron emission. The free-free and IR emissions
are  correlated on all scales studied ($0.4\leq a \leq10$\,kpc) in
both M~33 and M~31. The synchrotron--IR correlations, however,
{   are different} in M~31 and M~33, on small scales ($<$~1\,kpc) as
well as on larger scales. In M~31 only  little correlation
exists on scales $<1$\,kpc, but in M~33 the synchrotron--IR
correlation is reasonable on small scales.
We propose that the differences can be explained by the difference
in  diffusion length of CREs caused by the different magnetic
field properties of M~31 and M~33.

{  Combining our results with data on NGC\,6946, LMC and M~51,
we find evidence that the smallest scale, on which the synchrotron--IR
correlation exists, depends on the ratio of the
ordered-to-turbulent magnetic field strength. We show that this is consistent with
diffusion of CREs in the disk until they escape into the halo. The confinement
times of the CREs in the disks must be about the same. In our sample,
no correlation exists between $l_{\rm dif}$ and $\Sigma_{SFR}$.}

Our results show the fundamental importance of CRE propagation and
magnetic field structure for the radio--IR correlation within a
galaxy, and offer a new method to measure the diffusion length
$l_{\rm dif}$ of CREs in galaxies.

\section{Outlook}
\label{outlook}
Whether the propagation length of CREs in galaxies indeed
increases with increasing $B_{\rm ord}/B_{\rm tur}$, as our results
suggest, should be determined from the synchrotron--IR
cross-correlation as a function of scale for a large number of galaxies
with different degrees of field ordering and field strengths. For each
galaxy a realistic separation of free-free and synchrotron emission is
required \citep{Tabatabaei_2_07}, like we did for M~31 and M~33 (see
Sect.~\ref{free}).

{  Future observations should be performed at different frequencies.
At frequencies lower than 1\,GHz the propagation length of CREs should not
change much if the CRE confinement time is independent of CRE energy.
On the other hand, the lifetime of CREs in galaxies radiating at frequencies
higher than about 5\,GHz is expected to be dominated by synchrotron loss,
so that the dependence of propagation length on the field strength changes
from Eq.~(\ref{dif2}) to Eq.~(\ref{dif1}). The difference between the
propagation lengths at two widely separated frequencies will 
allow us to measure the diffusion coefficient and to provide constraints for models of
cosmic-ray propagation.}

\acknowledgement FST acknowledges the support by the DFG via the
grant TA 801/1-1. RB acknowledges the support by DFG FOR1254.
We thank Annie Hughes, Brent Groves,  and the anonymous referee
{  for discussions and helpful comments. Discussions with Andreas Shalchi,
Gianfranco Brunetti and Torsten En{\ss}lin
on CRE diffusion improved our insight into this process. }

\bibliography{s.bib}

\end{document}